\documentclass[aps,eqsecnum,preprint,floats,epsf,epsfig,nofootinbib,letter]{revtex4}
\usepackage{epsfig}
\usepackage{multirow}
\usepackage{subfigure}

\begin{document}
\def\be{\begin{eqnarray}}
\def\en{\end{eqnarray}}
\def\up{\uparrow}
\def\dw{\downarrow}
\def\non{\nonumber}
\def\la{\langle}
\def\ra{\rangle}
\def\T{{\cal T}}
\def\O{{\cal O}}
\def\B{{\cal B}}
\def\ep{\varepsilon}
\def\hep{\hat{\varepsilon}}
\def\ek{{\vec{\ep}_\perp\cdot\vec{k}_\perp}}
\def\epp{{\vec{\ep}_\perp\cdot\vec{P}_\perp}}
\def\kp{{\vec{k}_\perp\cdot\vec{P}_\perp}}
\def\lsim{ {\ \lower-1.2pt\vbox{\hbox{\rlap{$<$}\lower5pt\vbox{\hbox{$\sim$}
}}}\ } }
\def\gsim{ {\ \lower-1.2pt\vbox{\hbox{\rlap{$>$}\lower5pt\vbox{\hbox{$\sim$}
}}}\ } }
\def\dk{\partial\!\cdot\!K}
\def\pr{{\sl Phys. Rev.}~}
\def\prl{{\sl Phys. Rev. Lett.}~}
\def\pl{{\sl Phys. Lett.}~}
\def\np{{\sl Nucl. Phys.}~}
\def\zp{{\sl Z. Phys.}~}

\font\el=cmbx10 scaled \magstep2{\obeylines\hfill March, 2019}

\vskip 1.5 cm

\centerline{\large\bf Lifetimes of Doubly Heavy Baryons ${\cal B}_{bb}$ and ${\cal B}_{bc}$}
\bigskip
\centerline{\bf Hai-Yang Cheng$^a$, Fanrong Xu$^b$}
\medskip
\centerline{$^a$Institute of Physics, Academia Sinica}
\centerline{Taipei, Taiwan 115, Republic of China}
\medskip
\medskip
\centerline{$^b$ Department of Physics, Jinan University}
\centerline{Guangzhou 510632, People's Republic of China}
\bigskip
\bigskip
\bigskip
\bigskip
\bigskip
\centerline{\bf Abstract}
\bigskip
\small
Lifetimes of the doubly heavy baryons ${\cal B}_{bb}$ and ${\cal B}_{bc}$  are analyzed within the framework of the heavy quark expansion (HQE). Lifetime differences arise from the spectator effects such as $W$-exchange and Pauli interference.  For doubly bottom baryons, the lifetime pattern is  $\tau(\Omega_{bb}^-)\sim \tau(\Xi_{bb}^{-})>\tau(\Xi_{bb}^0)$.
The $\Xi_{bb}^{0}$ baryon is
shortest-lived owing to the $W$-exchange contribution, while $\Xi_{bb}^{-}$ and $\Omega_{bb}^{-}$ have similar lifetimes as they both receive contributions from destructive Pauli interference. We find the lifetime ratio $\tau(\Xi_{bb}^{-})/\tau(\Xi_{bb}^0)=1.26$\,.
The large $W$-exchange contribution to $\Xi_{bc}^0$ through the subprocess $cd\to us\to cd$ and the sizable destructive Pauli interference contribution to $\Xi_{bc}^+$ imply a substantial lifetime difference between $\Xi_{bc}^+$ and $\Xi_{bc}^0$.
In the presence of subleading $1/m_c$ and $1/m_b$ corrections to the spectator effects, we find that $\tau(\Omega_{bc}^0)$ becomes longest-lived. This is because $\Gamma^{\rm int}_+$ and $\Gamma^{\rm semi}$ for $\Omega_{bc}^0$ are subject to large cancellation between dimension-6 and -7 operators.  This implies that the subleading corrections are too large to justify the validity of the HQE. Demanding that $\Gamma^{cs}_{{\rm int+}}(\Omega_{bc}^0)$, $\Gamma^{{\rm SL},cs}_{\rm int}(\Omega_{bc}^0)$ be positive and $\Gamma^{cu}_{{\rm int-}}(\Xi^+_{bc})$ be negative, we conjecture that $1.68\times 10^{-13}s<\tau(\Omega_{bc}^0)< 3.70\times 10^{-13}s$ ,  $4.09\times 10^{-13}s<\tau(\Xi_{bc}^+)< 6.07\times 10^{-13}s$ and $0.93\times 10^{-13}s<\tau(\Xi_{bc}^0)< 1.18\times 10^{-13}s$.
Hence, the lifetime hierarchy of ${\cal B}_{bc}$ baryons is expected to be $\tau(\Xi_{bc}^{+})>\tau(\Omega_{bc}^0)>\tau(\Xi_{bc}^0)$.

\pagebreak

\section{Introduction}
After the discovery of the doubly charmed baryon $\Xi_{cc}^{++}$ in the $\Lambda_c^+K^-\pi^+\pi^+$ mass spectrum \cite{LHCb:Xiccpp}, LHCb proceeded to measure its lifetime   \cite{LHCb:tauXiccpp}
\be \label{eq:LHCbtauXiccpp}
\tau(\Xi_{cc}^{++})=
(2.56^{+0.24}_{-0.22}\pm0.14)\times 10^{-13}s.
\en
The theoretical predictions of doubly charmed baryon lifetimes in the literature \cite{Kiselev:1999,Kiselev:2002,Guberina,Chang,Karliner:2014,Cheng:doubly,Berezhnoy} listed in Table \ref{tab:lifetimes_dc} spread a large range, especially for $\Xi_{cc}^{++}$.  It appears that the early predictions of  $\tau(\Xi_{cc}^{++})$ were too large compared to experiment. The lifetime pattern is expected to be $\tau(\Xi_{cc}^{++})>\tau(\Omega_{cc}^+)>\tau(\Xi_{cc}^+)$.

\begin{table}[b]
\caption{Lifetimes of doubly charmed baryons
in units of $10^{-13}s$. The results of \cite{Berezhnoy} are based on the calculation of using $m_c=1.73\pm0.07$ GeV and $m_s=0.35\pm0.20$ GeV from a fit to the LHCb measurement of $\tau(\Xi_{cc}^{++}).$
} \label{tab:lifetimes_dc}
\begin{center}
\begin{tabular}{l c c c } \hline \hline
& ~~$\Xi_{cc}^{++}$~~ & ~~$\Xi_{cc}^{+}$~~ & ~~$\Omega_{cc}^{+}$~~~~~\\
\hline
~~Kiselev et al. ('99) \cite{Kiselev:1999}~~ &  ~~~$4.3\pm1.1$~~~ & ~~~$1.1\pm0.3$~~~ &  \\
~~Guberina et al. ('99) \cite{Guberina}~~ & 15.5 & 2.2 & 2.5 \\
~~Kiselev et al. ('02) \cite{Kiselev:2002}~~ & $4.6\pm0.5$ & $1.6\pm0.5$ & $2.7\pm0.6$ \\
~~Chang et al. ('04) \cite{Chang}~~  & 6.7 & 2.5 & 2.1 \\
~~Karliner, Rosner ('14) \cite{Karliner:2014}~~ & 1.85 & 0.53 & \\
~~Cheng, Shi ('18) \cite{Cheng:doubly} & 2.98 & 0.44 & 0.75$\sim$1.80\\
~~Berezhnoy et al. ('18) \cite{Berezhnoy} & $2.6\pm0.3$ & $1.4\pm0.1$ & $1.8\pm0.2$ \\
\hline
~~Expt. \cite{LHCb:tauXiccpp} & $2.56^{+0.28}_{-0.26}$ & & \\
\hline
\end{tabular}
\end{center}
\end{table}

\begin{table}[t]
\caption{Predicted lifetimes of doubly bottom and charm-bottom baryons
in units of $10^{-13}s$. } \label{tab:lifetimes_bc}
\begin{center}
\begin{tabular}{c|c c c c l  } \hline \hline
 & ~Likhoded et al. & ~Kiselev  et al.~  & ~Kiselev et al. & ~Karliner et al.~~ & Berezhnoy et al.  \\
 &  \cite{Likhoded} & \cite{Kiselev:1999kh} & \cite{Kiselev:2002} & \cite{Karliner:2014} & ~~~~~\cite{Berezhnoy}  \\
\hline
 ~~$\Xi_{bb}^{0}$~ & 7.9 &  & 7.9 & 3.7 & $5.2\pm0.095$  \\
 ~~$\Xi_{bb}^{-}$~ & 8.0 &  & 8.0 & 3.7 & $5.3\pm0.096$ \\
 ~~$\Omega_{bb}^{-}$~ & 8.0 & & 8.0 & & $5.3\pm0.093$  \\ \hline
~~$\Xi_{bc}^{+}$~ & 2.8 &$3.3\pm0.8$   & $3.0\pm0.4$ & 2.44 & $2.4\pm0.2$  \\
 ~~$\Xi_{bc}^{0}$~ & 2.6 & $2.8\pm0.7$  & $2.7\pm0.3$ & 0.93 & $2.2\pm0.18$   \\
 ~~$\Omega_{bc}^{0}$~ & 2.1 &  & $2.2\pm0.4$ & & $1.8\pm0.088$   \\ \hline
 \hline
\end{tabular}
\end{center}
\end{table}

\begin{figure}[t]
\begin{center}
\includegraphics[height=30mm]{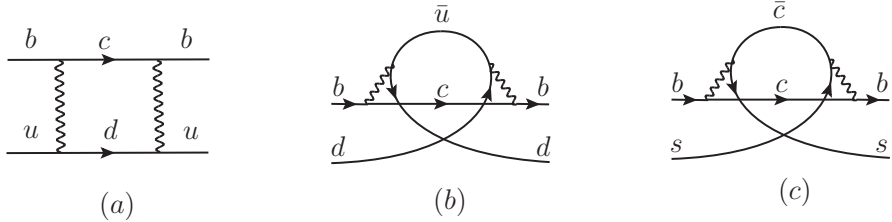}
\caption{Spectator effects in the nonleptonic decays of the doubly bottom baryons:  (a) $W$-exchange in $\Xi_{bb}^{0}$ decay and destructive Pauli interference in (b) $\Xi_{bb}^-$ and (c) $\Omega_{bb}^-$ decays.}
\label{fig:spectatorbb}
\end{center}
\end{figure}

\begin{figure}[t]
\begin{center}
\subfigure[]{
\includegraphics[height=21mm]{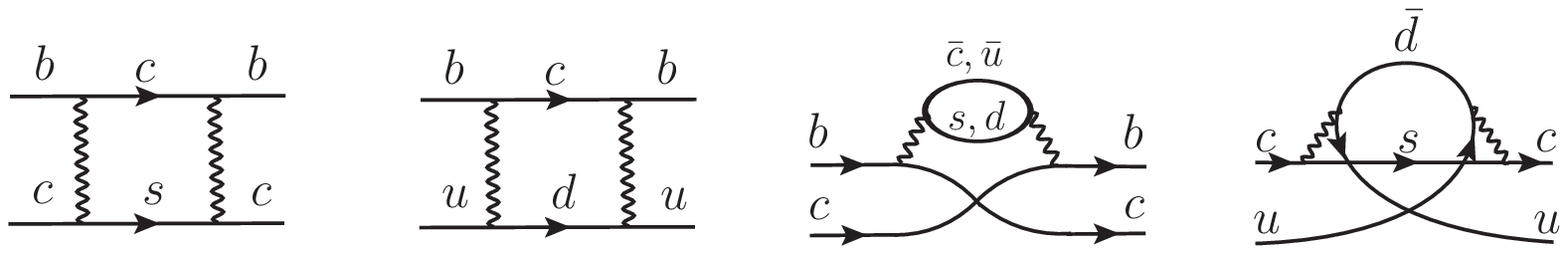}}
\subfigure[]{
\vspace{0.0cm}
\includegraphics[height=20mm]{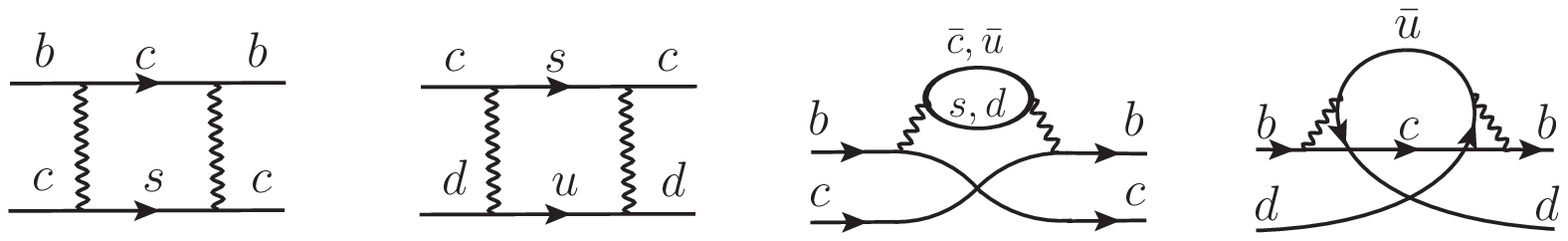}}
\subfigure[]{
\vspace{0.0cm}
\includegraphics[height=20mm]{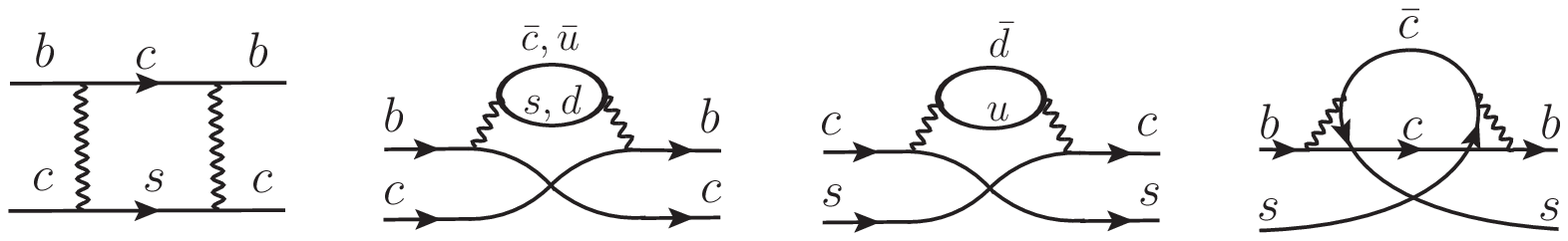}
}
\caption{Spectator effects in the nonleptonic weak decays of the charm-bottom baryons: (a) $\Xi_{bc}^{+}$, (b) $\Xi_{bc}^0$, and (c) $\Omega_{bc}^0$. In (a) and (b) there are two $W$-exchange diagrams, one constructive and one destructive Pauli interferences. In (c) there are one $W$-exchange, two constructive and one destructive Pauli interferences.}
\label{fig:spectatorbc}
\end{center}
\end{figure}

In this work, we would like to generalize our previous study of doubly charmed baryon lifetimes \cite{Cheng:doubly} to the doubly bottom  baryons $\B_{bb}$ and charm-bottom baryons $\B_{bc}$. Some predictions available in the literature are shown in Table \ref{tab:lifetimes_bc}. It is well known that the lifetime differences stem mainly from the spectator effects such as weak annihilation and Pauli interference. Spectator effects are depicted in Figs. \ref{fig:spectatorbb} and \ref{fig:spectatorbc} for $\B_{bb}$ and $\B_{bc}$ baryons, respectively. Calculations in \cite{Kiselev:2002,Karliner:2014,Berezhnoy,Likhoded} indicate that the lifetimes of $\Xi_{bb}^0$ and $\Xi_{bb}^-$ are close to each other. However, we see from Fig. \ref{fig:spectatorbb} that $\Xi_{bb}^0$ has a positive contribution from the $W$-exchange box diagram, while both $\Xi_{bb}^-$ and $\Omega_{bb}^-$ receive destructive Pauli interference contributions. Hence, it is anticipated that $\tau(\Omega_{bb}^-)\sim \tau(\Xi_{bb}^-)> \tau(\Xi_{bb}^0)$. We are going to show in this study that this is indeed the case. Likewise, a large $W$-exchange contribution to $\Xi_{bc}^0$ through the subprocess $cd\to us\to cd$ and a large destructive Pauli interference contribution to $\Xi_{bc}^+$ (see Fig. \ref{fig:spectatorbc}) will imply a substantial lifetime difference between $\Xi_{bc}^+$ and $\Xi_{bc}^0$, which will be  checked  in this work.

The study of $\B_{bc}$ lifetimes is more complicated than the $\B_{bb}$ case for several reasons. First, besides the spectator effects due to each heavy quark $b$ or $c$, there also exist $W$-exchange and Pauli interference in which both $b$ and $c$ quarks get involved.
Second, care must be taken when considering the heavy quark expansion (HQE) for the charm quark.
It is known that the HQE in $1/m_b$ works well for bottom hadrons \cite{Cheng:2018}.
On the contrary, the HQE to $1/m_c^3$ fails to give a satisfactory description of the lifetimes of both charmed mesons and charmed baryons \cite{Cheng:2018}.  Since the charm quark is not heavy,
it is thus natural to consider the effects arising from the next order $1/m_c$ expansion. This calls for the subleading $1/m_Q$ corrections to the spectator effects.
It turns out that although the relevant dimension-7 spectator effects are in the right direction for explaining the large lifetime ratio of the charmed baryons such as $\tau(\Xi_c^+)/\tau(\Lambda_c^+)$, the destructive $1/m_c$ corrections to the lifetime of $\Omega_c^0$ are too large to justify the use of the HQE, namely, the predicted Pauli interference and semileptonic rates for the $\Omega_c^0$ become negative, which certainly do not make sense.
Demanding these rates to be positive for a sensible HQE, it has been conjectured in \cite{Cheng:2018} that the $\Omega_c^0$ lifetime lies in the range of $(2.3\sim3.2)\times 10^{-13}s$. This is indeed consistent with the new measurement of the $\Omega_c^0$ lifetime by LHCb \cite{LHCb:Omegac} and the new lifetime pattern
$\tau(\Xi_c^+)>\tau(\Omega_c^0)>\tau(\Lambda_c^+)>\tau(\Xi_c^0)$, contrary to the hierarchy $\tau(\Xi_c^+)>\tau(\Lambda_c^+)>\tau(\Xi_c^0)>\tau(\Omega_c^0)$ given in the PDG \cite{PDG}.
This indicates that
the $\Omega_c^0$, which is naively expected to be shortest-lived in the charmed baryon system owing to the large constructive Pauli interference, could live longer than the $\Lambda_c^+$  due to the suppression from next order $1/m_c$ corrections arising from dimension-7 four-quark operators.
By the same token, this effect should be also taken into account in both $\B_{bb}$ and $\B_{bc}$ systems.

For the doubly heavy baryons $\B_{bc}$, we shall follow the traditional convention of unprimed states $\Xi_{bc}$ and $\Omega_{bc}$ with the $bc$ diquark being of the axial-vector type ($S_{bc}=1$) and primed states $\Xi'_{bc}$ and $\Omega'_{bc}$ with the scalar type of the heavy diquark ($S_{bc}=0$). Theoretical calculations imply that $\B'_{bc}$ is heavier than $\B_{bc}$.
\footnote{Almost all the calculations available in the literature lead to $m_{\Xi'_{bc}}>m_{\Xi_{bc}}$ (see, e.g. Fig. 18 of \cite{Meinel}) except the QCD sum-rule calculation of \cite{Tang} and the lattice QCD calculation of \cite{Meinel}. Note that the definition of primed and unprimed $\B_{bc}$ states in \cite{Meinel} is opposite to the conventional one.}
Hence, the primed $\B_{bc}'$ baryons are dominated by the electromagnetic decay $\B'_{bc}\to\B_{bc}\gamma$.

In this work we shall study the lifetimes of doubly heavy baryons within the framework of the HQE. It is organized as follows. In Sec.~II we give the
general HQE expressions for inclusive nonleptonic and
semileptonic widths. A special attention is paid to the doubly heavy baryon matrix elements of dimension-3 and -5 operators which are somewhat different from the ones of singly heavy baryons. We then proceed to discuss the relevant dimension-6 and -7 four-quark operators. Evaluation of doubly heavy baryon matrix elements and numerical results are presented in Sec. III. Conclusions are given in Sec.~IV.

\section{Theoretical framework}
Under the heavy quark expansion,
the inclusive
nonleptonic decay rate of a doubly heavy baryon $\B_{QQ'}$ containing two heavy quarks $QQ'$
is given by \cite{Bigi92,BS93}
\be
\Gamma(\B_{QQ'})={1\over 2m_{\B_{QQ'}}}{\rm Im}\,\la \B_{QQ'}|T|\B_{QQ'}\ra={1\over 2m_{\B_{QQ'}}}\la \B_{QQ'}|\int d^4x\,T[{\cal L}^\dagger_W(x){\cal L}_W(0)]|\B_{QQ'}\ra,
\en
in analog to the case of a singly heavy baryon $\B_Q$.
Through the use of the operator product expansion, the transition operator $T$ can be expressed in terms of local quark operators
\be \label{eq:HQE}
{\rm Im}\,T={G_F^2m_Q^5\over 192\pi^3}\,\xi\,\left(c_{3,Q}\bar QQ+{c_{5,Q}\over m_Q^2}\bar Q\sigma\cdot G Q+{c_{6,Q}\over m_Q^3}T_6+ {c_{7,Q}\over m_Q^4}T_7+\cdots\right),
\en
where $\xi$ is the relevant CKM matrix element,
the dimension-6 $T_6$ consists of the four-quark operators $(\bar Q\Gamma q)(\bar q\Gamma Q)$ with $\Gamma$ representing a combination of the Lorentz and color matrices, while a subset of dimension-7 $T_7$ is governed by the four-quark operators containing derivative insertions. Hence,
\be \label{eq:NLrate}
\Gamma(\B_{bc}) &=& {G_F^2m_b^5\over 192\pi^3}\,\xi\,{1\over 2m_{\B_{bc}}}
\Bigg\{ c_{3,b}\la \B_{bc}|\bar bb|\B_{bc}\ra+ {c_{5,b}\over m_b^2} \la\B_{bc}|\bar b \sigma\cdot Gb|\B_{bc}\ra \non \\ &+& {c_{6,b} \over m_b^3} \la \B_{bc}|T_6|\B_{bc}\ra+{c_{7,b} \over m_b^4} \la \B_{bc}|T_7|\B_{bc}\ra+\cdots\Bigg\}  \non \\
&+& {G_F^2m_c^5\over 192\pi^3}\,\xi\,{1\over 2m_{\B_{bc}}}
\Bigg\{ c_{3,c}\la \B_{bc}|\bar cc|\B_{bc}\ra+ {c_{5,c}\over m_c^2} \la\B_{bc}|\bar c \sigma\cdot Gc|\B_{bc}\ra \non \\ &+& {c_{6,c} \over m_c^3} \la \B_{bc}|T_6|\B_{bc}\ra+{c_{7,c} \over m_c^4} \la \B_{bc}|T_7|\B_{bc}\ra+\cdots\Bigg\}.
\en

In the following we shall discuss the contributions from dimension-3, -5, -6 and -7 operators separately.

\subsection{Dimension-3 and -5 operators}
In heavy quark effective theory (HQET), the dimension-3 operator $\bar QQ$ in the rest frame has the expression
\be
\bar QQ=\bar Q\gamma_0 Q-{\bar Q(i\vec{D})^2Q\over 2m_Q^2}+{\bar Q\sigma\cdot G Q\over 4m_Q^2}+{\cal O}\left({1\over m_Q^3}\right),
\en
with the normalization
\be
{\la \B_{QQ'}|\bar Q\gamma_0 Q|\B_{QQ'}\ra \over 2m_{\B_{QQ'}}}=1.
\en
Hence,
\be \label{eq:dim3me}
{\la \B_{bc}|\bar cc|\B_{bc}\ra\over 2m_{\B_{bc}}}=1-{\mu_{\pi,c}^2\over 2m_c^2}+{\mu_{G,c}^2\over 2m_c^2}+{\cal O}\left({1\over m_c^3}\right),
\en
where
\be \label{eq:lambda12}
&& \mu_{\pi,c}^2\equiv {1\over 2m_{\B_{bc}}}\la \B_{bc}|\bar c(i\vec{D})^2c|\B_{bc}\ra=-{1\over 2m_{\B_{bc}}}\la \B_{bc}|\bar c(i{D_\perp})^2c|\B_{bc}\ra=-\lambda_1^c,   \non \\
&& \mu_{G,c}^2\equiv{1\over 2m_{\B_{bc}}}\la \B_{bc}|\bar c{1\over 2}\sigma\cdot
Gc|\B_{bc}\ra=d_H\lambda_2^c.
\en
The non-perturbative parameters $\lambda_1$ and $\lambda_2$ are independent of $m_Q$ and have the same values for all particles in a given spin-flavor multiplet.

We first consider the non-perturbative parameter $\mu_\pi^2$. In general, $\mu_{\pi,Q}^2=\la p^2\ra=\la m_Q^2v_Q^2\ra$. The average kinetic energy of the diquark $bc$ and the light quark $q$ is $T={1\over 2}m_{di}v_{di}^2+{1\over 2}m_qv_q^2$, where $m_{di}$ ($m_q$) is the mass of the diquark (light quark). This together with the momentum conservation $m_{di}v_{di}=m_qv_q$ leads to
\be
v_{di}^2={2m_q T\over (m_b+m_c)(m_b+m_c+m_q)}.
\en
As shown in \cite{Kiselev:1999}, the average kinetic energy $T'$ of heavy quarks inside the diquark given by ${1\over 2}m_b(\tilde v_b^2+\tilde v_c^2)$ is equal to $T/2$ due to the color wave function of the diquark. Hence, the average velocity  $\tilde v_{b}$ of the heavy quark $b$ inside the diquark is $\tilde v_b^2=m_cT/[m_b(m_b+m_c)]$, where we have applied the momentum conservation $m_b\tilde v_b=m_c\tilde v_c$. As a result, the average velocity $v_Q$ of the heavy quark inside the baryon $\B_{bc}$ is \cite{Kiselev:2002}
\be
v_b^2 &\approx& \tilde v_b^2+v_{di}^2={m_cT\over m_b(m_b+m_c)}+{2m_q T\over (m_b+m_c)(m_b+m_c+m_q)}, \non \\
v_c^2 &\approx& \tilde v_c^2+v_{di}^2={m_bT\over m_c(m_b+m_c)}+{2m_q T\over (m_b+m_c)(m_b+m_c+m_q)}.
\en
Hence,
\be
\mu_{\pi,b}^2(\B_{bc}) &\simeq& m_b^2\left( {m_c T\over m_b(m_b+m_c)}+{2m_q T\over (m_b+m_c)(m_b+m_c+m_q)}\right), \non \\
\mu_{\pi,c}^2(\B_{bc}) &\simeq& m_c^2\left( {m_b T\over m_c(m_b+m_c)}+{2m_q T\over (m_b+m_c)(m_b+m_c+m_q)}\right).
\en

We next turn to the parameter $\mu_G^2$.  For the doubly heavy baryon $\B_{QQ'}$, if the heavy diquark acts as a point-like constitute, its mass is  of the form
\be \label{eq:mHQQ}
m_{\B_{bc}}=\,m_b+m_c+\bar\Lambda_{\B_{bc}}+{\mu_{\pi,b}^2\over 2m_b}+{\mu_{\pi,c}^2\over 2m_c}-{\mu_{G,b}^2\over 2m_b}-{\mu_{G,c}^2\over 2m_c}+{\cal O}\left({1\over m_Q^2}\right).
\en
There are two distinct chromomagnetic fields inside the $\B_{bc}$:
one is  the chromomagnetic field produced by the light quark and the other
by the heavy quark.
To proceed,  let us consider a simple quark model of De R\'ujula {\it et al.} \cite{DeRujula}
\be
M_{\rm baryon} &=& M_0+\cdots+{16\over 9}\pi\alpha_s\sum_{i>j}{\vec{S}_i\cdot\vec{S}_j\over m_im_j}|\psi(0)|^2, \non \\
M_{\rm meson} &=& M_0+\cdots+{32\over 9}\pi\alpha_s {\vec{S}_1\cdot\vec{S}_2\over m_1m_2}|\psi(0)|^2.
\en
It is well known that the fine structure constant is $-{4\over 3}\alpha_s$ for $\bar qq$ pairs in a meson and $-{2\over 3}\alpha_s$ for $qq$ pairs in a baryon \cite{DeRujula}. This is because the $\bar qq$ pair in a meson must be a color-singlet, while the $qq$ pair in a baryon is in color antitriplet state.
The mass of the doubly heavy baryon $\B_{bc}$ is given by
\be
m_{\B_{bc}}=m_b+m_c+\cdots+{16\over 9}\pi\alpha_s\left[\left( {\vec{S}_{b}\cdot\vec{S}_q \over m_bm_q }+{\vec{S}_{c}\cdot\vec{S}_q \over m_cm_q }\right)|\psi^{q,bc}(0)|^2+{\vec{S}_b\cdot\vec{S}_c\over m_bm_c}|\phi^{bc}(0)|^2\right],
\en
where  $\psi^{q,bc}(0)$ is the light quark wave function at the origin of the $bc$ diquark and $\phi^{bc}(0)$ is the diquark wave function at the origin.

The matrix elements of spin-spin interactions are given by
\be
\la \B_{bc}|\vec{S}_{b}\cdot\vec{S}_q|\B_{bc}\ra
 ={1\over 2}[S_{bq}(S_{bq}+1)-S_b(S_b+1)-S_q(S_q+1)], \non \\
 \la \B_{bc}|\vec{S}_{c}\cdot\vec{S}_q|\B_{bc}\ra
 ={1\over 2}[S_{cq}(S_{cq}+1)-S_c(S_c+1)-S_q(S_q+1)],
\en
where $S_{bq}$ ($S_{cq}$) is the spin of the diquark $bq$ ($cq$).
To evaluate the diquark spins $S_{bq}$ and $S_{cq}$, we need to change the basis from $|S,S_{bc}\ra$ to $|S,S_{bq}\ra$ or $|S,S_{cq}\ra$ with $S$ being the spin of the heavy baryon
\be
|S,S_{bc}\ra &=& \sum_{S_{bq}}(-1)^{(S+S_b+S_c+S_q)}\sqrt{(2S_{bq}+1)(2S_{bc}+1)}\left\{
\begin{array}{ccc}
    S_q  & S_b &  S_{bq} \\
    S_c & S & S_{bc} \\
   \end{array} \right\}|S,S_{bq}\ra, \non \\
   |S,S_{bc}\ra &=& \sum_{S_{cq}}(-1)^{(S+S_b+S_c+S_q)}\sqrt{(2S_{cq}+1)(2S_{bc}+1)}\left\{
\begin{array}{ccc}
    S_q  & S_c &  S_{cq} \\
    S_b & S & S_{bc} \\
   \end{array} \right\}|S,S_{cq}\ra.
\en
We find,
\be \label{eq:massXicc}
m_{\Xi_{bc}} &=& m_b+m_c+\cdots+{16\over 9}\pi\alpha_s\left( -{1\over 2}{m_b+m_c\over m_bm_cm_q} |\psi^{q,bc}(0)|^2+{1\over 4m_bm_c}|\phi^{bc}(0)|^2\right), \non \\
m_{\Xi_{bc}^*} &=& m_b+m_c+\cdots+{16\over 9}\pi\alpha_s\left( {1\over 4}{m_b+m_c\over m_bm_cm_q} |\psi^{q,bc}(0)|^2+{1\over 4m_bm_c}|\phi^{bc}(0)|^2\right).
\en
The term proportional to $|\psi^{q,bc}(0)|^2$ can be expressed in terms of the hyperfine mass splitting of $\Xi_{bc}$:
\be \label{eq:hyperfine}
m_{\Xi_{bc}^*}-m_{\Xi_{bc}}={4\over 3}\pi\alpha_s{m_b+m_c\over m_bm_cm_q}|\psi^{q,bc}(0)|^2.
\en
Identifying the last two terms in the parentheses of Eq. (\ref{eq:massXicc}) with $-\mu_{G,b}^2/(2m_b)-\mu_{G,c}^2/(2m_c)$,
we obtain
\be
\mu_{G,b(c)}^2(\Xi_{bc})={2\over 3}(m_{\Xi_{bc}^*}-m_{\Xi_{bc}})m_{b(c)}-{4\over 9}\pi\alpha_s{|\phi^{bc}(0)|^2\over m_{c(b)}}.
\en

It has been known that HQET is not the appropriate effective field theory for hadrons with more than one heavy quark.  For a singly heavy hadron, the heavy quark kinetic energy is neglected as it occurs as a small $1/m_Q$ correction. For a bound state containing two or more heavy quarks, the heavy quark kinetic energy is very important and cannot be treated as a perturbation. The appropriate theory for dealing such a system is non-relativistic QCD (NRQCD),
in which one has
\be
\bar Qg_s\sigma\cdot GQ=-2\psi_Q^\dagger g_s\vec{\sigma}\cdot\vec{B}\psi_Q-{1\over m_Q}\psi^\dagger_Q g_s\vec{D}\cdot \vec{E}\psi_Q+\cdots
\en
in terms of the two-spinor $\psi_Q$. According to the counting rule, the Darwin term for the interaction with the chromoelectric field is of the same order of magnitude as the chromomagnetic term \cite{Beneke:Bc}.  Hence, we get an additional contribution to $\mu_G^2$
\be \label{eq:muG}
\mu_{G,b}^2={2\over 3}(m_{\Xi_{bc}^*}-m_{\Xi_{bc}})m_b-{1\over 9}g_s^2{|\phi^{bc}(0)|^2\over m_c}-{1\over 6}g^2_s{|\phi^{bc}(0)|^2\over m_b}, \non \\
\mu_{G,c}^2={2\over 3}(m_{\Xi_{bc}^*}-m_{\Xi_{bc}})m_c-{1\over 9}g_s^2{|\phi^{bc}(0)|^2\over m_b}-{1\over 6}g^2_s{|\phi^{bc}(0)|^2\over m_c}.
\en
The last term in $\mu_{G,b}^2$ and $\mu_{G,c}^2$ can be obtained by using the equation of motion for the chromoelectric field.
It follows that
\be \label{eq:m.e.ofcc}
{\la \Xi_{bc}|\bar cc|\Xi_{bc}\ra\over 2m_{\Xi_{bc}}}=1-{1\over 2}v_c^2+{1\over 3}{ m_{\Xi_{bc}^*}-m_{\Xi_{bc}}\over m_c} -{1\over 18}g_s^2{|\phi^{bc}(0)|^2\over m_bm_c^2}-{1\over 12}g_s^2{|\phi^{bc}(0)|^2\over m_c^3}.
\en
This expression is different from the one given in \cite{Kiselev:2002}
\be
{\la \Xi_{bc}|\bar cc|\Xi_{bc}\ra\over 2m_{\Xi_{bc}}}=1-{1\over 2}v_c^2+{1\over 3}g_s^2{|\phi^{bc}(0)|^2\over m_bm_c^2}-{1\over 6}g_s^2{|\phi^{bc}(0)|^2\over m_c^3}.
\en
We notice that the third term in Eq. (\ref{eq:m.e.ofcc}) is absent in the above expression.
This is because the authors of \cite{Kiselev:2002} considered the charm-bottom baryon $\Xi'_{bc}$ with the scalar $bc$ diquark so that $S_{bc}=0$. It is straightforward to show that $\la\Xi'_{bc}|\vec{S}_b\cdot\vec{S}_q|\Xi'_{bc}\ra=\la\Xi'_{bc}|\vec{S}_c\cdot\vec{S}_q|\Xi'_{bc}\ra=0$. Hence, the chromomagnetic interaction of the heavy quark with the light quark  does not contribute to $\Xi'_{bc}$. However, since $\Xi'_{bc}$ is heavier than $\Xi_{bc}$,
it is dominated by the electromagnetic decay $\Xi'_{bc}\to\Xi_{bc}\gamma$.

The nonleptonic and semiletponic decay rates of the $\B_{bc}$ baryons are then given by
\be \label{eq:dec}
\Gamma^{\rm dec}(\B_{bc}) &=& {G_F^2m_c^5\over 192\pi^3}\,\xi\,
\Bigg\{ c_{3,c}^{\rm NL} \Big[1-{\mu_{\pi,c}^2\over 2m_c^2}+{\mu_{G,c}^2\over 2m_c^2}\Big]+ 2c_{5,c}^{\rm NL}{\mu_{G,c}^2\over m_c^2}\Bigg\}  \non \\
&+& {G_F^2m_b^5\over 192\pi^3}\,\xi\,
\Bigg\{ c_{3,b}^{\rm NL} \Big[1-{\mu_{\pi,b}^2\over 2m_b^2}+{\mu_{G,b}^2\over 2m_b^2}\Big]+ 2c_{5,b}^{\rm NL}{\mu_{G,b}^2\over m_b^2}\Bigg\},
\en
and
\be \label{eq:SLrate}
\Gamma^{\rm SL}(\B_{bc}) &=& {G_F^2m_c^5\over 192\pi^3}\,\xi\,
\Bigg\{ c_{3,c}^{\rm SL} \Big[1-{\mu_{\pi,c}^2\over 2m_c^2}+{\mu_{G,c}^2\over 2m_c^2}\Big]+ 2c_{5,c}^{\rm SL}{\mu^2_{G,c}\over m_c^2} \Bigg\}  \non \\
&+& {G_F^2m_b^5\over 192\pi^3}\,\xi\,
\Bigg\{ c_{3,b}^{\rm SL} \Big[1-{\mu_{\pi,b}^2\over 2m_b^2}+{\mu_{G,b}^2\over 2m_b^2}\Big]+ 2c_{5,b}^{\rm SL}{\mu^2_{G,b}\over m_b^2} \Bigg\},
\en
where the expressions of the coefficients $c_{3,b(c)}$ and $c_{5,b(c)}$ can be found, for example, in \cite{Cheng:2018}. For doubly bottom baryons $\B_{bb}$, the expressions of $\Gamma^{\rm dec}(\B_{bb})$ and $\Gamma^{\rm SL}(\B_{bb})$ are the same as Eqs. (\ref{eq:dec}) and (\ref{eq:SLrate}), respectively, except that the charm quark is replaced by the bottom quark.

\subsection{Dimension-6 operators}

 Defining
\be
\T_6={G_F^2m_Q^2\over 192\pi^3}\xi\,c_{6,Q}^{\rm NL}\,T_6,
\en
the dimension-6 four-quark operators in Eq. (\ref{eq:NLrate}) for spectator effects in inclusive decays of doubly charmed baryons $\B_{bc}$ are given by (only Cabibbo-allowed decays being listed here)
\cite{Kiselev:1999kh}
\be  \label{eq:T6baryon}
\T_{6,ann,bc}^{\B_{bc}} &=& {G^2_Fm_b^2\over 2\pi}\,|V_{cb}|^2\,\left(1+\sqrt{x_b}\,\right)^2\left(1-x_{b+}\right)^2\Big\{
(c_1^2+c_2^2)(\bar bb)(\bar cc)+2c_1c_2(\bar bc)(\bar cb)\Big\},
 \non \\
\T_{6,ann,bu}^{\B_{bc}}
 &=& {G^2_Fm_b^2\over 2\pi}\,|V_{cb}|^2\,(1-x_b)^2\Big\{
(c_1^2+c_2^2)(\bar bb)(\bar uu)+2c_1c_2(\bar bu)(\bar ub)\Big\},
 \non \\
\T_{6,ann,cd}^{\B_{bc}}  &=& {G^2_Fm_c^2\over 2\pi}\,|V_{cs}|^2\,\left(1-x_c\right)^2\Big\{
(c_1^2+c_2^2)(\bar cc)(\bar dd)+2c_1c_2(\bar cd)(\bar dc)\Big\},
 \non \\
\T_{6,int-,cu}^{\B_{bc}} &=& -{G_F^2m_c^2\over 6\pi}\,|V_{cs}|^2 (1-x_c)^2\Bigg\{ c_1^2\left[ (1+
{1\over 2}x_c)(\bar cc)(\bar uu)-(1+2x_c)\bar c^\alpha(1-\gamma_5)u^\beta
\bar u^\beta(1+\gamma_5)c^\alpha\right]   \non \\
&+& (2c_1c_2+N_cc_2^2)\left[ (1+{1\over 2}x_c)(\bar cu)(\bar uc)-
(1+2x_c)\bar c(1-\gamma_5)u\bar u(1+\gamma_5)c\right]\Bigg\},   \non \\
\T_{6,int-,bd}^{\B_{bc}} &=& -{G_F^2m_b^2\over 6\pi}\,|V_{cb}|^2 (1-x_b)^2\Bigg\{ c_1^2\left[ (1+
{1\over 2}x_b)(\bar bb)(\bar dd)-(1+2x_b)\bar b^\alpha(1-\gamma_5)d^\beta
\bar d^\beta(1+\gamma_5)b^\alpha\right]   \non \\
&+& (2c_1c_2+N_cc_2^2)\left[ (1+{1\over 2}x_b)(\bar bd)(\bar db)-
(1+2x_b)\bar b(1-\gamma_5)d\bar d(1+\gamma_5)b\right]\Bigg\},    \\
\T_{6,int-,bs}^{\B_{bc}} &=& -{G_F^2m_b^2\over 6\pi}\,|V_{cb}|^2 \sqrt{1-4x_b}\Bigg\{ c_1^2\left[ (1-
x_b)(\bar bb)(\bar ss)-(1+2x_b)\bar b^\alpha(1-\gamma_5)s^\beta
\bar s^\beta(1+\gamma_5)b^\alpha\right]   \non \\
&+& (2c_1c_2+N_cc_2^2)\left[ (1-x_b)(\bar bs)(\bar sb)-
(1+2x_b)\bar b(1-\gamma_5)s\bar s(1+\gamma_5)b\right]\Bigg\},  \non  \\
\T_{6,int+,cs}^{\B_{bc}} &=& -{G_F^2m_c^2\over 6\pi}\,|V_{cs}|^2\left(1-\sqrt{x_c}\,\right)^2\Bigg\{
c_2^2\left[(\bar cc)(\bar ss)-\bar c^\alpha(1-\gamma_5)s^\beta
\bar s^\beta(1+\gamma_5)c^\alpha\right]   \non \\
&+& (2c_1c_2+N_cc_1^2)\Big[ (\bar cs)(\bar sc)-
\bar c(1-\gamma_5)s\bar s(1+\gamma_5)c\Big] \Bigg\}, \non \\
\T_{6,int+,bc}^{\B_{bc}} &=& -{G_F^2m_b^2\over 6\pi}\,|V_{cb}V_{cs}|^2\left(1-\sqrt{x_b}\,\right)^2\left(1-x_{b-}\right)^2 \non \\
&\times & \Bigg\{
c_2^2\left[(1+{1\over 2}x_{b-})(\bar bb)(\bar cc)-(1+2x_{b-})\bar b^\alpha(1-\gamma_5)c^\beta
\bar c^\beta(1+\gamma_5)b^\alpha\right]   \non \\
&+& (2c_1c_2+N_cc_1^2)\Big[(1+{1\over 2}x_{b-}) (\bar bc)(\bar cb)-
(1+2x_{b-})\bar b(1-\gamma_5)c\bar c(1+\gamma_5)b\Big] \Bigg\} \non \\
&-& {G_F^2m_b^2\over 6\pi}\,|V_{cb}V_{ud}|^2\left(1-\sqrt{x_b}\,\right)^2\Big\{ \cdots {\rm with~}x_{b-}\to 0\Big\}, \non
\en
where $(\bar q_1q_2)\equiv \bar q_1\gamma_\mu(1-\gamma_5)q_2$, and $\alpha,~\beta$
are color indices, $x_c=m_s^2/m_c^2$, $x_b=m_c^2/m_b^2$ and $x_{b\pm}=m_c^2/(m_b\pm m_c)^2$.

Spectator effects in the weak decays of the doubly charmed baryons $\Xi_{bc}^{+}$, $\Xi_{bc}^{0}$ and $\Omega_{bc}^0$ are depicted in Fig. \ref{fig:spectatorbc}. The first amplitude $\T_{6,ann,bc}^{\B_{bc}}$ in (\ref{eq:T6baryon}) corresponds to the first $W$-exchange diagram in Fig. \ref{fig:spectatorbc}(a), Fig. \ref{fig:spectatorbc}(b) and Fig. \ref{fig:spectatorbc}(c), which is common to all $\B_{bc}$ baryons. Similarly, the amplitude $\T_{6,ann,bu}^{\B_{bc}}$ corresponds to the second $W$-exchange diagram in Fig. \ref{fig:spectatorbc}(a). The amplitude $\T_{6,int+,bc}^{\B_{cc}}$ arises from the constructive Pauli interference of the $c$ quark produced in the $b$ quark decay with the $c$ quark in the wave function of $\B_{bc}$. It corresponds to the third diagram in Fig. \ref{fig:spectatorbc}(a) and Fig. \ref{fig:spectatorbc}(b), and the second diagram in Fig. \ref{fig:spectatorbc}(c).  The term $\T_{6,int-,cu}^{\B_{bc}}$ is due to the destructive interference of the $u$ quark and it occurs in the fourth diagram in Fig. \ref{fig:spectatorbc}(a).

For inclusive semileptonic decays of $\B_{bc}$ baryons, there is a spectator effect originating from the constructive Pauli interference of the $c$ or $s$
quark \cite{Voloshin}; that is, the $c$ ($s$) quark produced in $b\to c\ell^-\bar\nu_\ell$ ($c\to s\ell^+\nu_\ell$) has an interference with the $c$ ($s$) quark in the wave function of $\B_{bc}$ ($\Omega_{bc}^0$). This amounts to replacing the loop quarks $s\bar c$ and $d\bar u$ ($u\bar d$) in the second (third) diagram of Fig. \ref{fig:spectatorbc}(c) by $\ell^-\bar \nu_\ell$ ($\ell^+\nu_\ell$).
It is now ready to deduce this term from $\T_{6,int+,cs}^{\B_{bc}}$ and $\T_{6,int+,bc}^{\B_{bc}}$ in Eq. (\ref{eq:T6baryon})  by putting $c_1=1$, $c_2=0$ and $N_c=1$:
\be \label{eq:SL6}
\T_{6,int,bc}^{\B_{bc},\rm SL} &=& -{G_F^2m_b^2\over 6\pi}\,|V_{cb}|^2(1-\sqrt{x_b}\,)^2
 \Big[(1+{1\over 2}x_{b\ell}) (\bar bc)(\bar cb)-
(1+2x_{b\ell})\bar b(1-\gamma_5)c\bar c(1+\gamma_5)b\Big], \non \\
\T_{6,int,cs}^{\B_{bc},\rm SL} &=&
 -{G_F^2m_c^2\over 6\pi}|V_{cs}|^2\,(1-\sqrt{x_c}\,)^2
\Big[(1+{1\over 2}x_{c\ell}) (\bar cs)(\bar sc)-
(1+2x_{c\ell})\bar c(1-\gamma_5)s\bar s(1+\gamma_5)c\Big], \non \\
\en
where $x_{c\ell}=m_\ell^2/m_c^2$ and $x_{b\ell}=m_\ell^2/m_b^2$.

For doubly bottom baryons $\B_{bb}$, the expressions of $\T_{6,ann,bu}^{\B_{bb}}$, $\T_{6,int-,bd}^{\B_{bb}}$ and $\T_{6,int-,bs}^{\B_{bb}}$ (see Fig. \ref{fig:spectatorbb}) have the same expressions as $\T_{6,ann,bu}^{\B_{bc}}$, $\T_{6,int-,bd}^{\B_{bc}}$ and $\T_{6,int-,bs}^{\B_{bc}}$, respectively, in Eq. (\ref{eq:T6baryon}). However, there is no additional spectator effect in semileptonic decays of $\B_{bb}$ baryons.

\subsection{Dimension-7 operators}
To the order of $1/m_Q^4$ in the heavy quark expansion in Eq. (\ref{eq:NLrate}), we need to consider dimension-7 operators. For our purposes, we shall focus on the $1/m_Q$
corrections to the spectator effects discussed in the last subsection and neglect the operators with gluon fields.
Dimension-7 terms are either the four-quark operators times the spectator quark mass or the four-quark operators with one or two additional derivatives \cite{Gabbiani:2003pq,Gabbiani:2004tp}.

We obtain \cite{Cheng:2018}
\be \label{eq:T7baryon}
\T_{7,ann,bc}^{\B_{bc}} &=& {G^2_Fm_b^2\over 2\pi}\,|V_{cb}|^2\,\left(1+\sqrt{x_b}\right)^2(1-x_{b+}\,)^2\Bigg\{ 2c_1c_2\Big[2(1+x_b)P_3^{bc}+(1-x_b)P^{bc}_5\Big] \non \\
&+& (c_1^2+c_2^2)\left[2(1+x_b)\tilde P_3^{bc}+(1-x_b)\tilde P^{bc}_5\right]\Bigg\},
\non\\
\T_{7,ann,bu}^{\B_{bc}} &=& {G^2_Fm_b^2\over 2\pi}\,|V_{cb}|^2\,(1-x_b)^2\Bigg\{ 2c_1c_2\Big[2(1+x_b)P_3^{bu}+(1-x_b)P^{bu}_5\Big] \non \\
&+& (c_1^2+c_2^2)\left[2(1+x_b)\tilde P_3^{bu}+(1-x_b)\tilde P^{bu}_5\right]\Bigg\},
\non\\
\T_{7,ann,cd}^{\B_{bc}} &=& {G^2_Fm_c^2\over 2\pi}\,|V_{cs}|^2\,(1-x_c)^2\Bigg\{ 2c_1c_2\Big[2(1+x_c)P_3^{cd}+(1-x_c)P^{cd}_5\Big] \non \\
&+& (c_1^2+c_2^2)\left[2(1+x_c)\tilde P_3^{cd}+(1-x_c)\tilde P^{cd}_5\right]\Bigg\},
\non\\
\T_{7,int,cu}^{\B_{bc}} &=& {G_F^2m_c^2\over 6\pi}\,|V_{cs}|^2(1-x_c)^2\Bigg\{ \Big(2c_1c_2+N_cc_2^2\Big)\Big[-(1-x_c)(1+2x_c)(P_1^{cu}+P_2^{cu}) \non \\
&+& 2(1+x_c+x_c^2)P_3^{cu}
-12x_c^2P_4^{cu}-(1-x_c)(1+{1\over 2}x_c)P_5^{cu}+(1-x_c)(1+2x_c)P_6^{cu}\Big] \non \\
 &+& c_1^2\Big[ -(1-x_c)(1+2x_c)(\tilde P_1^{cu}+\tilde P_2^{cu})
+ 2(1+x_c+x_c^2)\tilde P_3^{cu}-12x_c^2\tilde P_4^{cu} \non \\
&-& (1-x_c)(1+{1\over 2}x_c)\tilde P_5^{cu}+(1-x_c)(1+2x_c)\tilde P_6^{cu}\Big]
\Bigg\},   \\
\T_{7,int,bd}^{\B_{bc}} &=& \T_{7,int,cu}^{\B_{bc}}(c\to b, u\to d, V_{cs}\to V_{cb}), \non \\
\T_{7,int,bs}^{\B_{bc}} &=& \T_{7,int,bd}^{\B_{bc}}(d\to s), \non \\
\T_{7,int,cs}^{\B_{bc}} &=& {G_F^2m_c^2\over 6\pi}\,|V_{cs}|^2\left(1-\sqrt{x_c}\,\right)^2\Bigg\{ \left(2c_1c_2+N_cc_1^2\right)\Big[-P_1^{cs}-P_2^{cs}
+ 2P_3^{cs} -P_5^{cs}+P_6^{cs}\Big] \non \\
 &+& c_2^2\Big[ -\tilde P_1^{cs}-\tilde P_2^{cs}
+ 2\tilde P_3^{cs} - \tilde P_5^{cs}+\tilde P_6^{cs}\Big]
\Bigg\}, \non \\
\T_{7,int,bc}^{\B_{bc}} &=& {G_F^2m_b^2\over 6\pi}\,|V_{cb}|^2\left(1-\sqrt{x_b}\,\right)^2\Bigg\{ \left(2c_1c_2+N_cc_1^2\right)\Big[-P_1^{bc}-P_2^{bc}
+ 2P_3^{bc} -P_5^{bc}+P_6^{bc}\Big] \non \\
 &+& c_2^2\Big[ -\tilde P_1^{bc}-\tilde P_2^{bc}
+ 2\tilde P_3^{bc} - \tilde P_5^{bc}+\tilde P_6^{bc}\Big]
\Bigg\}, \non
\en
where dimension-7 four-quark operators are defined by \cite{Lenz:D}
\footnote{The $T_7$ term in the HQE is suppressed by a factor of $1/m_Q$ relative to $T_6$ (see Eq. (\ref{eq:HQE})). However,  this suppression factor is absorbed in the definition of $P_i^{Qq}$ for later convenience.
We shall see below that the hadronic matrix elements of dimension-7 operators are suppressed relative to that of dimension-6 ones by order $m_q/m_Q$.
}
\be
&& P_1^{Qq}={m_q\over m_Q}\bar Q(1-\gamma_5)q\bar q(1-\gamma_5)Q, \qquad\qquad\qquad
~~P_2^{Qq}={m_q\over m_Q}\bar Q(1+\gamma_5)q\bar q(1+\gamma_5)Q, \non \\
&& P_3^{Qq}={1\over m_Q^2}\bar Q \stackrel{\leftarrow}{D}_\rho\gamma_\mu(1-\gamma_5)D^\rho q\bar q\gamma^\mu(1-\gamma_5)Q, \quad
~P_4^{Qq}={1\over m_Q^2}\bar Q \stackrel{\leftarrow}{D}_\rho(1-\gamma_5)D^\rho q\bar q(1+\gamma_5)Q, \non \\
&& P_5^{Qq}={1\over m_Q}\bar Q \gamma_\mu(1-\gamma_5)q\bar q\gamma^\mu(1-\gamma_5)(iD\!\!\!\!/)Q, \quad\quad
~~P_6^{Qq}={1\over m_Q}\bar Q (1-\gamma_5)q\bar q(1+\gamma_5)(iD\!\!\!\!/)Q, \non \\
\en
and $\tilde P_i$ denotes the color-rearranged operator that follows from the expression of $P_i$ by interchanging the color indices of the $Q_i$ and $q_j$ Dirac spinors, for example, $\tilde P_1^{Qq}={m_q\over m_Q}\bar Q_i(1-\gamma_5)q_j\bar q_j(1-\gamma_5)Q_i$.

For doubly bottom baryons $\B_{bb}$, the spectator effects $\T_{7,ann,bu}^{\B_{bb}}$, $\T_{7,int,bd}^{\B_{bb}}$ and $\T_{7,int,bs}^{\B_{bb}}$ have the same expressions as   $\T_{7,ann,bu}^{\B_{bc}}$, $\T_{7,int,bd}^{\B_{bc}}$ and $\T_{7,int,bs}^{\B_{bc}}$ in Eq. (\ref{eq:T7baryon}), respectively.

As for the dimension-7 contributions to semileptonic decays, it can be obtained  from $\T_{7,int}^{\B_{bc},bc}$ and $\T_{7,int}^{\B_{bc},cs }$by setting $c_1=1$, $c_2=0$ and $N_c=1$. Taking into account the lepton mass corrections, it reads \cite{Cheng:2018}
\be  \label{eq:SL7}
\T_{7,int}^{{\rm SL},bc} &=& {G_F^2m_b^2\over 6\pi}\,|V_{cb}|^2\left(1-\sqrt{x_b}\,\right)^2\Big[-(1-x_{b\ell})^2(1+2x_{b\ell})(P_1^{bc}+P_2^{bc}) \non \\
&+& 2(1-x_{b\ell})(1+x_{b\ell}+x_{b\ell}^2)P_3^{bc}
 -12x_{b\ell}^2(1-x_{b\ell})P_4^{bc}\Big], \non \\
\T_{7,int}^{{\rm SL},cs} &=& {G_F^2m_c^2\over 6\pi}\,|V_{cs}|^2\left(1-\sqrt{x_c}\,\right)^2\Big[-(1-x_{c\ell})^2(1+2x_{c\ell})(P_1^{cs}+P_2^{cs}) \non \\
&+& 2(1-x_{c\ell})(1+x_{c\ell}+x_{c\ell}^2)P_3^{cs}
 -12x_{c\ell}^2(1-x_{c\ell})P_4^{cs}\Big],
\en
where $x_{Q\ell}=(m_\ell/m_Q)^2$.

\section{Lifetimes of doubly heavy baryons $\B_{bc}$ and $\B_{bb}$ }

\subsection{Baryon matrix elements}

The spectator effects in inclusive decays of the charm-bottom baryons $\B_{bc}$ arising from dimension-6 and dimension-7 operators are given by Eqs. (\ref{eq:T6baryon}), (\ref{eq:SL6}), (\ref{eq:T7baryon}) and (\ref{eq:SL7}), respectively.
We shall rely on the quark model to evaluate the baryon matrix elements of four-quark operators. Since the heavy $bc$ diquark of $\B_{bc}$ is of the axial-vector type, its flavor-spin wave function is given by
\be
\B_{bc}={1\over 6}\left(2b^\up c^\up q^\dw-b^\up c^\dw q^\up-b^\dw c^\up q^\up+2c^\up b^\up q^\dw-c^\up b^\dw q^\up-c^\dw b^\up q^\up+ (13)+(23)\right).
\en
In the nonrelativistic quark model we have (see Appendix B of \cite{Cheng:2018} for the detail)
\be
\la \Xi_{bc}|(\bar cc)(\bar qq)|\Xi_{bc}\ra = 6 \, m_{\Xi_{bc}}|\psi^{q,bc}(0)|^2,
\en
where we have taken into account the normalization of the matrix element $\la \B_{QQ'}|Qv\!\!\!/ Q|\B_{QQ'}\ra$ or $\la \B_{QQ'}|Q'v\!\!\!/ Q'|\B_{QQ'}\ra$ to $2m_{B_{QQ'}}$.
The relevant $\B_{bc}$ baryon matrix elements of dimension-6 operators are
\be \label{eq:m.e.dim6}
&& \la \B_{bc}|(\bar Qq)(\bar qQ)|\B_{bc}\ra = -6 \, m_{\B_{bc}}|\psi^{q,bc}(0)|^2,  \non \\
&& \la \B_{bc}|(\bar QQ)(\bar qq)|\B_{bc}\ra = 6 \, m_{\B_{bc}}|\psi^{q,bc}(0)|^2\tilde B, \non \\
&& \la \B_{bc}|\bar Q(1-\gamma_5)q\bar q(1+\gamma_5)Q|\B_{bc}\ra = - m_{\B_{bc}}|\psi^{q,bc}(0)|^2, \\
&& \la \B_{bc}|\bar Q^\alpha(1-\gamma_5)q^\beta\bar q^\beta(1+\gamma_5)Q^\alpha|
\B_{bc}\ra =  m_{\B_{bc}}|\psi^{q,bc}(0)|^2\tilde B, \non
\en
with $Q=b,c$. The parameter $\tilde B$ is defined by
\be \label{eq:tildeB}
\la \B_{bc}|(\bar QQ)(\bar qq)|\B_{bc}\ra=-\tilde B \la \B_{bc}|(\bar Qq)(\bar qQ)|\B_{bc}\ra.
\en
Since the color wavefunction for a baryon is totally antisymmetric,
the matrix element of $(\bar QQ)(\bar qq)$ is the same as that of $(\bar Qq)
(\bar qQ)$ except for a sign difference. That is, $\tilde B=1$ under the valence-quark approximation.

For the $\B_{bc}$ matrix elements of four-quark operators involved both $b$  and $c$ quarks, we obtain
\be \label{eq:m.e.dim6bc}
&& \la \B_{bc}|(\bar bc)(\bar cb)|\B_{bc}\ra = 0,  \non \\
&& \la \B_{bc}|(\bar bb)(\bar cc)|\B_{bc}\ra = 0, \non \\
&& \la \B_{bc}|\bar b(1-\gamma_5)c\bar c(1+\gamma_5)b|\B_{bc}\ra = 2\,m_{\B_{bc}}|\phi^{bc}(0)|^2, \\
&& \la \B_{bc}|\bar b^\alpha(1-\gamma_5)c^\beta\bar c^\beta(1+\gamma_5)b^\alpha|
\B_{bc}\ra = -2\,m_{\B_{bc}}|\phi^{bc}(0)|^2\tilde B. \non
\en
It should be remarked that the $\B_{bc}$ matrix elements of the four-quark operators $(\bar bc)(\bar cb)$ and $(\bar bb)(\bar cc)$ vanish in the nonrelativistic quark model but not in the MIT bag model. However, for the reason of consistency, we will stick to the former model.
Note that our expressions of the $\B_{bc}$ matrix elements of dimension-6 operators Eqs. (\ref{eq:m.e.dim6}) and (\ref{eq:m.e.dim6bc}) are different from that given in \cite{Kiselev:1999kh} and \cite{Kiselev:2002} in which the spin of the $bc$ diquark is treated to be zero.

Likewise,  the $\B_{bc}$ matrix elements of the dimension-7 operators $P_i^{Qq}$ read
\be \label{eq:m.e.dim7}
&& \la \B_{bc}|P_1^{Qq}|\B_{bc}\ra=\la \B_{bc}|P_2^{Qq}|\B_{bc}\ra= {3\over 2}\,m_{\B_{bc}}|\psi^{q,bc}(0)|^2\left( {m^2_{\B_{bc}}-m_{\{bc\}}^2\over m_Q(m_b+m_c)}\right)\eta^q_{1,2},  \non \\
&& \la \B_{bc}|P_3^{Qq}|\B_{bc}\ra= 3\la \B_{bc}|P_4^{Qq}|\B_{bc}\ra=-3\,m_{\B_{bc}}|\psi^{q,bc}(0)|^2\left( {m^2_{\B_{bc}}-m_{\{bc\}}^2\over m_Q(m_b+m_c)}\right)\eta^q_{3,4},
\en
where the parameters $\eta_i^q$ are expected to be of order unity, and $m_{\{bc\}}$  is the mass of the axial-vector $bc$ diquark. In the derivation of Eq. (\ref{eq:m.e.dim7}) we have applied the relations $p_b\approx m_b v$ and $p_c\approx m_c v$ as the $bc$ system has been treated as a diquark. It is then straightforward to show that
\be
{p_b\cdot p_q\over m_b^2}\approx {1\over 2}{m_{\B_{bc}}^2-m_{\{bc\}}^2\over m_b(m_b+m_c)}, \qquad  {p_c\cdot p_q\over m_c^2}\approx {1\over 2}{m_{\B_{bc}}^2-m_{\{bc\}}^2\over m_c(m_b+m_c) }.
\en
Therefore, the matrix elements of dimension-7 operators are suppressed by a factor of $m_q/m_Q$ relative to that of dimension-6 ones.
Matrix elements of the dimension-7 operators $P_i^{bc}$ read
\be \label{eq:m.e.bcdim7}
&& \la \B_{bc}|P_1^{bc}|\B_{bc}\ra=\la \B_{bc}|P_2^{bc}|\B_{bc}\ra=2 \,m_{\B_{bc}}\left({m_c\over m_b}\right)|\phi^{bc}(0)|^2\eta^q_{1,2},  \non \\
&& \la \B_{bc}|P_3^{bc}|\B_{bc}\ra= 0, \non \\
&& \la \B_{bc}|P_4^{bc}|\B_{bc}\ra=-4\,m_{\B_{bc}}\left({m_c\over m_b}\right)|\phi^{bc}(0)|^2\eta^q_{4}.
\en
They are suppressed by a factor of $m_c/m_b$ relative to the matrix elements of dimension-6 operators.
For the matrix elements of the operators $\tilde P_i^{Qq}$,  we introduce a parameter $\tilde\beta_i^q$ in analog to Eq. (\ref{eq:tildeB})
\be \label{eq:beta}
\la \B_{bc}|\tilde P_i^{Qq}|\B_{bc}\ra=-\tilde \beta_i^q \la \B_{bc}|P_i^{Qq}|\B_{bc}\ra,
\en
so that $\tilde\beta_i^q=1$ under the valence quark approximation.

The flavor-spin wave function of $\B_{bb}$ is given by
\be
\B_{bb}={1\over \sqrt{18}}\left(2b^\up b^\up q^\dw-b^\up b^\dw q^\up-b^\dw b^\up q^\up+ (13)+(23)\right).
\en
Then the relevant $\B_{bb}$ matrix elements have the expressions 
\be \label{eq:m.e.bbdim6}
&& \la \B_{bb}|(\bar bq)(\bar qb)|\B_{bb}\ra = -12m_{_{\B_{bb}}}|\psi^{q,bb}(0)|^2,  \non \\
&& \la \B_{bb}|(\bar bb)(\bar qq)|\B_{bb}\ra = 12m_{_{\B_{bb}}}|\psi^{q,bb}(0)|^2  \tilde B, \non \\
&& \la \B_{bb}|\bar b(1-\gamma_5)q\bar q(1+\gamma_5)b|\B_{bb}\ra = -2m_{_{\B_{bb}}}|\psi^{q,bb}(0)|^2, \\
&& \la \B_{bb}|\bar b^\alpha(1-\gamma_5)q^\beta\bar q^\beta(1+\gamma_5)b^\alpha|
\B_{bb}\ra = 2m_{_{\B_{bb}}}|\psi^{q,bb}(0)|^2\tilde B, \non
\en
and
\be \label{eq:m.e.dim7bb}
&& \la \B_{bb}|P_1^{bq}|\B_{bb}\ra=\la \B_{bb}|P_2^{bq}|\B_{bb}\ra={3\over 2} m_{_{\B_{bb}}}|\psi^{q,bb}(0)|^2 \left( {m^2_{\B_{bb}}-m_{\{bb\}}^2\over m_b^2}\right)\eta^q_{1,2},  \non \\
&& \la \B_{bb}|P_3^{bq}|\B_{bb}\ra= 6\la \B_{bb}|P_4^{bq}|\B_{bb}\ra=-3 m_{_{\B_{bb}}}|\psi^{q,bb}(0)|^2\left( {m^2_{\B_{bb}}-m^2_{\{bb\}}\over m_b^2}\right)\eta^q_{3,4}.
\en
In numerical calculations, we shall take $m_{\{bc\}}$ to be 6526 MeV and $m_{\{bb\}}$ 9778 MeV obtained from the relativistic quark model \cite{Ebert:2005}.

We are ready to evaluate the spectator effects in $\B_{bc}$ decays given by
\be
\Gamma^{\rm spec}(\B_{bc})={\la \B_{bc}|\T_6+\T_7|\B_{bc}\ra\over 2m_{\B_{bc}}}.
\en
The results are
\be \label{eq:Spectorbc}
\Gamma^{\B_{bc}}_{{\rm ann},bc} &=& -{G_F^2m_b^2\over 2\pi}\,|V_{cb}|^2\,(1+\sqrt{x_b})^2 (1+x_b)(1-x_{b+})^2
\Big(\tilde \beta(c_1^2+c_2^2)-2c_1c_2\Big)\eta
\left({m_c\over m_b}\right)\left|\phi^{bc}(0)\right|^2   \non \\
\Gamma^{\B_{bc}}_{{\rm ann},bu} &=& {3\over 2}{G_F^2m_b^2\over \pi}\,|V_{cb}|^2\,(1-x_{b})^2 \left|\psi^{q,bc}(0)\right|^2\Bigg\{
\Big(\tilde B(c_1^2+c_2^2)-2c_1c_2\Big)   \non \\
&+& (1+x_b)
\Big(\tilde \beta(c_1^2+c_2^2)-2c_1c_2\Big)\eta
\left( {m^2_{\B_{bc}}-m^2_{\{bc\}}\over m_b(m_b+m_c) } \right)\Bigg\},
   \non \\
\Gamma^{\B_{bc}}_{{\rm ann},cd} &=& {3\over 2}{G_F^2m_c^2\over \pi}\,|V_{cs}|^2\,(1-x_{c})^2 \left|\psi^{q,bc}(0)\right|^2\Bigg\{
\Big(\tilde B(c_1^2+c_2^2)-2c_1c_2\Big)   \non \\
&+& (1+x_c)\Big(\tilde \beta(c_1^2+c_2^2)-2c_1c_2\Big)\eta
\left( {m^2_{\B_{bc}}-m^2_{\{bc\}}\over m_c(m_b+m_c) } \right)\Bigg\},
\non \\
\Gamma^{\B_{bc}}_{{\rm int}_-,cu} &=& -{G_F^2m_c^2\over 12\pi}\,|V_{cs}|^2\, (1-x_c)^2 \left|\psi^{q,bc}(0)\right|^2
\Bigg\{ \Big(\tilde B c_1^2-2c_1c_2-N_cc_2^2\Big)(5+x_c)
\non \\
&-&  9\Big(\tilde \beta c_1^2-2c_1c_2-N_cc_2^2\Big)(1+x_c-{2\over 3}x_c^2)\eta\left( {m^2_{\B_{bc}}-m^2_{\{bc\}}\over m_c(m_b+m_c)} \right)\Bigg\},   \non \\
\Gamma^{\B_{bc}}_{{\rm int}_-,bd} &=& -{G_F^2m_b^2\over 12\pi}\,|V_{cb}|^2\, (1-x_b)^2 \left|\psi^{q,bc}(0)\right|^2
\Bigg\{ \Big(\tilde B c_1^2-2c_1c_2-N_cc_2^2\Big)(5+x_b)
\non \\
&-&  9\Big(\tilde \beta c_1^2-2c_1c_2-N_cc_2^2\Big)(1+x_b-{2\over 3}x_b^2)\eta\left( {m^2_{\B_{bc}}-m^2_{\{bc\}}\over m_b(m_b+m_c)} \right)\Bigg\},   \non \\
\Gamma^{\B_{bc}}_{{\rm int}_-,bs} &=& -{G_F^2m_b^2\over 12\pi}\,|V_{cb}|^2\, \sqrt{1-4x_b} \left|\psi^{s,bc}(0)\right|^2
\Bigg\{ \Big(\tilde B c_1^2-2c_1c_2-N_cc_2^2\Big)(5+x_b)
\non \\
&-&  9\Big(\tilde \beta c_1^2-2c_1c_2-N_cc_2^2\Big)(1+x_b-{2\over 3}x_b^2)\eta\left( {m^2_{\B_{bc}}-m^2_{\{bc\}}\over m_b(m_b+m_c)} \right)\Bigg\},   \\
\Gamma^{\B_{bc}}_{{\rm int}_+,cs} &=& {G_F^2m_c^2\over 12\pi}\,|V_{cs}|^2\, (1-\sqrt{x_c})^2\left|\psi^{s,bc}(0)\right|^2
\Bigg\{ 5\Big(2c_1c_2+N_cc_1^2-\tilde B c_2^2\Big)
\non \\
&-&  9\Big(2c_1c_2+N_cc_1^2-\tilde \beta c_2^2 \Big)\eta\left( {m^2_{\B_{bc}}-m^2_{\{bc\}}\over m_c(m_b+m_c)} \right)\Bigg\},  \non  \\
\Gamma^{\B_{bc}}_{{\rm int}_+,bc} &=& {G_F^2m_b^2\over 6\pi}\,|V_{cb}V_{cs}|^2\, (1-\sqrt{x_b})^2\left|\phi^{bc}(0)\right|^2
\Bigg\{ \Big(2c_1c_2+N_cc_1^2-\tilde B c_2^2\Big)(1-x_{b-})^2(1+2x_{b-})
\non \\
&-&  2\Big(2c_1c_2+N_cc_1^2-\tilde \beta c_2^2 \Big)\eta\left({m_c\over m_b}\right)\Bigg\}  \non\\
&+& {G_F^2m_b^2\over 6\pi}\,|V_{cb}V_{ud}|^2\, (1-\sqrt{x_b})^2\left|\phi^{bc}(0)\right|^2\Big\{\cdots {\rm with~}x_{b-}\to 0\Big\}, \non
\en
and
\be \label{eq:SL67bc}
\Gamma^{{\rm SL},bc}_{int} &=& {G_F^2m_b^2\over 6\pi}|V_{cb}|^2\, (1-\sqrt{x_b})^2\left|\phi^{bc}(0)\right|^2 \left[1-2\left({m_c\over m_b} \right)\right], \non \\
\Gamma^{{\rm SL},cs}_{int}
&=& {G_F^2m_c^2\over 12\pi}|V_{cs}|^2(1-\sqrt{x_c})^2\left|\psi^{s,bc}(0)\right|^2  \left[ 5-9\left({m_{\B_{bc}}^2-m_{\{bc\}}^2\over m_c(m_b+m_c)}\right)\right].
\en
Except for the weak annihilation term, the expression of Pauli interference will be very lengthy if the hadronic parameters $\eta^q_i$ and $\tilde \beta^q_i$ are all treated to be different from each other. Since in realistic calculations we will set $\tilde \beta^q_i(\mu_h)=1$ under the valence quark approximation and put $\eta^q_i$ to unity, we shall assume for simplicity that $\eta^q_i=\eta$ and $\tilde \beta^q_i=\tilde\beta$.

For the Wilson coefficients in Eqs. (\ref{eq:Spectorbc}) and (\ref{eq:SL67bc}), we choose the scale $\mu$ to be $m_b$ ($m_c$) for the $b$ ($c$) quark decay and for the spectator effect involved the $b$ ($c$) quark. For example, we choose $\mu\approx m_b$ for the Wilson coefficients in $\Gamma^{\B_{bc}}_{{\rm int}_-,bd}$ and  $\mu\approx m_c$ for the Wilson coefficients in $\Gamma^{\B_{bc}}_{{\rm int}_+,cs}$. For the spectator effect involved both $b$ and $c$ quarks, for example $\Gamma^{\B_{bc}}_{{\rm int}_+,bc}$, we shall follow \cite{Beneke:Bc} to set $\mu=2m_r$ with $m_r$ being the reduced mass $m_bm_c/(m_b+m_c)$.

There are two quantaties in Eqs. (\ref{eq:Spectorbc}) and (\ref{eq:SL67bc}) which we need to know, namely $|\psi^{q,bc}(0)|$ and $|\phi^{bc}(0)|$.
From Eq. (\ref{eq:hyperfine}) we see that the wave function of the light quark at the origin of the $bc$ diquark $|\psi^{q,bc}(0)|$ is related to the hyperfine mass splitting of $\Xi_{bc}$.  To remove the dependence on the light quark mass $m_q$, we notice that the hyperfine mass splitting of $B$ mesons is given by
\be
m_{B^*}-m_B={32\over 9}\alpha_s\pi {|\psi^{b\bar q}_B(0)|^2\over m_bm_q}.
\en
Hence,
\be \label{eq:rinBbc}
|\psi^{q,bc}(0)|^2 &=&  {8\over 3}\,{\alpha_s(m_b)\over \alpha_s(2m_r)}\,{m_c\over m_b+m_c}\,{m_{\Xi_{bc}^*}-m_{\Xi_{bc}}\over m_{B^*}-m_B}|\psi_B^{b\bar q}(0)|^2\equiv r_{\Xi_{bc}}|\psi_B^{b\bar q}(0)|^2
\non \\
|\psi^{s,bc}(0)|^2 &=&  {8\over 3}\,{\alpha_s(m_b)\over \alpha_s(2m_r)}\,{m_c\over m_b+m_c}\,{m_{\Omega_{bc}^*}-m_{\Omega_{bc}}\over m_{B_s^*}-m_{B_s} }|\psi_{B_s}^{b\bar s}(0)|^2\equiv r_{\Omega_{bc}}|\psi_{B_s}^{b\bar s}(0)|^2
\en
where the $B$ meson wave functions at the origin squared are given by
\be
|\psi_B^{b\bar q}(0)|^2={1\over 12}f_B^2m_B, \qquad \quad |\psi_{B_s}^{b\bar s}(0)|^2={1\over 12}f_{B_s}^2m_{B_s}.
\en
As for the wave function of the diquark $QQ'$ at the origin, we shall use \cite{Baranov}
\be
|\phi^{cc}(0)|^2=0.039\,{\rm GeV}^3, \qquad |\phi^{bc}(0)|^2=0.065\,{\rm GeV}^3, \qquad
|\phi^{bb}(0)|^2=0.152\,{\rm GeV}^3.
\en

For the doubly bottom $\B_{bb}$ baryons, we have
\be
\Gamma^{\Xi_{bb}^0}_{{\rm ann},bu} &=& 3{G_F^2m_b^2\over \pi}\,|V_{cb}|^2\,(1-x_{b})^2 \left|\psi^{q,bb}(0)\right|^2\Bigg\{
\Big(\tilde B(c_1^2+c_2^2)-2c_1c_2\Big)   \non \\
&+& {1\over 2}(1+x_b)
\Big(\tilde \beta(c_1^2+c_2^2)-2c_1c_2\Big)\eta
\left( {m^2_{\Xi_{bb}}-m^2_{\{bb\}}\over m_b^2 } \right)\Bigg\},
\non \\
\Gamma^{\Xi_{bb}^-}_{{\rm int}_-,bd} &=& -{G_F^2m_b^2\over 6\pi}\,|V_{cb}|^2\, (1-x_b)^2 \left|\psi^{q,bb}(0)\right|^2
\Bigg\{ \Big(\tilde B c_1^2-2c_1c_2-N_cc_2^2\Big)(5+x_b)
\non \\
&-&  {9\over 2}\Big(\tilde \beta c_1^2-2c_1c_2-N_cc_2^2\Big)(1+x_b-{2\over 3}x_b^2)\eta\left( {m^2_{\Xi_{bb}}-m^2_{\{bb\}}\over m_b^2} \right)\Bigg\},    \\
\Gamma^{\Omega_{bb}^-}_{{\rm int}_-,bs} &=& -{G_F^2m_b^2\over 6\pi}\,|V_{cb}|^2\, (1-x_b)^2 \left|\psi^{s,bb}(0)\right|^2
\Bigg\{ \Big(\tilde B c_1^2-2c_1c_2-N_cc_2^2\Big)(5+x_b)
\non \\
&-&  {9\over 2}\Big(\tilde \beta c_1^2-2c_1c_2-N_cc_2^2\Big)(1+x_b-{2\over 3}x_b^2)\eta\left( {m^2_{\Omega_{bb}}-m^2_{\{bb\}}\over m_b^2} \right)\Bigg\},   \non
\en
where  the wave function of the light quark at the origin of the $bb$ diquark $|\psi^{q,bb}(0)|$ is given by
\be \label{eq:rinBbb}
|\psi^{q,bb}(0)|^2 &=&  {4\over 3}\,{m_{\Xi_{bb}^*}-m_{\Xi_{bb}}\over m_{B^*}-m_B}|\psi_B^{b\bar q}(0)|^2\equiv r_{\Xi_{bb}}|\psi_B^{b\bar q}(0)|^2, \non \\
|\psi^{s,bb}(0)|^2 &=&  {4\over 3}\,{m_{\Omega_{bb}^*}-m_{\Omega_{bb}}\over m_{B_s^*}-m_{B_s} }|\psi_{B_s}^{b\bar s}(0)|^2 \equiv r_{\Omega_{bb}}|\psi_{B_s}^{b\bar s}(0)|^2.
\en

\subsection{Numerical results}

To compute the decay widths of doubly heavy baryons,
we have to specify the values of $\tilde B$ and $r_{\B_{QQ'}}$. Since $\tilde B=1$ in the
valence-quark approximation and since the wavefunction squared ratio $r_{\B_{QQ'}}$ defined in Eq. (\ref{eq:rinBbc}) or (\ref{eq:rinBbb})
is evaluated using the quark model, it is reasonable to assume that the NQM
and the valence-quark approximation are most reliable when the baryon matrix
elements are evaluated at a typical hadronic scale $\mu_{\rm had}$. As
shown in \cite{Neubert97}, the parameters $\tilde B$ and $r$ renormalized
at two different scales are related via the renormalization group equation
to be
\be \label{eq:RGE1}
\tilde B(\mu)r(\mu) =\, \tilde B(\mu_{\rm had})r(\mu_{\rm had}),  \qquad \tilde B(\mu) =\, {\tilde{B}(\mu_{\rm had})\over \kappa+{1\over N_c}(\kappa
-1)\tilde B(\mu_{\rm had}) }\,,
\en
with
\be \label{eq:RGE2}
\kappa=\left({\alpha_s(\mu_{\rm had})\over \alpha_s(\mu)}\right)^{3N_c/2
\beta_0}=\sqrt{\alpha_s(\mu_{\rm had})\over \alpha_s(\mu)}
\en
and $\beta_0={11\over 3}N_c-{2\over 3}n_f$. The parameter $\kappa$ takes care of the evolution from $m_Q$ to the hadronic scale.
We consider the hadronic scale in the range of $\mu_{\rm had}\sim 0.65-1$ GeV. Taking  the  scale $\mu_{\rm had}=0.9$ GeV as an illustration, we obtain $\tilde{B}(\mu)=(0.75,0.67,0.57)\tilde B(\mu_{\rm
had})\simeq (0.75,0.67,0.57)$ and $r(\mu)\simeq (1.33,1.50,1.74)\,r(\mu_{\rm had})$.
The parameter $\tilde\beta$ is treated in a similar way.

We shall discuss the lifetimes of the doubly bottom baryons $\B_{bb}$ first for their simplicity. Following \cite{Cheng:2018}, we will use the kinetic $b$ quark mass $m_b=4.546$ GeV as the calculated inclusive semileptonic $B$ rate to the leading order (LO) using this kinetic mass is very close to  the experimental measurement \cite{Cheng:2018}.
For numerical calculations, we use the LO Wilson coefficients $c_1(\mu)=1.104$ and $c_2(\mu)=-0.243$ evaluated at the scale $\mu=4.546$ GeV, $m_{\Xi_{bb}^*}-m_{\Xi_{bb}}=35$ MeV and $m_{\Omega_{bb}^*}-m_{\Omega_{bb}}=30$ MeV from \cite{Ebert:2005}, the wave function $|\psi^{bb}(0)|^2=0.151\,{\rm GeV}^{3}$ \cite{Baranov} and the average kinetic energy $T=0.37$ GeV \cite{Kiselev:1999}. For the decay constants, we use $f_B=186$ MeV and $f_{B_s}=230$ MeV.
For the charm quark mass we use $m_c=1.56$ GeV fixed from the experimental values for $D^+$ and $D^0$ semileptonic widths \cite{Cheng:2018}.

\begin{table}[t]
\caption{Various contributions to the decay rates (in units of
$10^{-13}$ GeV) of doubly bottom baryons $\B_{bb}$ to order $1/m_b^4$ in the HQE with the hadronic scale $\mu_{\rm had}=0.825$ GeV.
}
\label{tab:lifetime_Bbb}
\begin{center}
\begin{tabular}{l c c c r c   c  } \hline \hline
 & $\Gamma^{\rm dec}$ & ~$\Gamma^{\rm ann}$ & $\Gamma^{\rm int}_-$  & ~ $\Gamma^{\rm semi}$~~ & $\Gamma^{\rm tot}$ & ~$\tau(10^{-13}s)$~   \\
\hline
 ~$\Xi_{bb}^{0}$ & ~~5.800 &  ~1.698 & & 2.090~~ & 9.587
 & ~ 6.87~  \\
 ~$\Xi_{bb}^-$ & ~~5.800 &  & ~$-0.283$~ & 2.090~~ & 7.607
 & ~ 8.65~    \\
 ~$\Omega_{bb}^-$ & ~~5.802 &  & $-0.312$ & 2.092~~ & $7.583$
 & ~ 8.68~    \\
\hline \hline
\end{tabular}
\end{center}
\end{table}

The total widths read
\be
\Gamma(\Xi_{bb}^0) &=& \Gamma^{\rm dec}(\Xi_{bb}^0)+\Gamma^{\rm SL}(\Xi_{bb}^0)+\Gamma^{\B_{bb}}_{{\rm ann},bu}, \non \\
\Gamma(\Xi_{bb}^-) &=& \Gamma^{\rm dec}(\Xi_{bb}^-)+\Gamma^{\rm SL}(\Xi_{bb}^-) +\Gamma^{\B_{bb}}_{{\rm int}_-,bd},  \\
\Gamma(\Omega_{bb}^-) &=& \Gamma^{\rm dec}(\Omega_{bb})+\Gamma^{\rm SL}(\Omega_{bb})+\Gamma^{\B_{bb}}_{{\rm int}_-,bs}. \non
\en
The results of calculations to order $1/m_b^4$ are shown in Table \ref{tab:lifetime_Bbb}.
The lifetime pattern is  $\tau(\Omega_{bb}^-)\sim \tau(\Xi_{bb}^{-})>\tau(\Xi_{bb}^0)$.
The $\Xi_{bb}^{0}$ baryon is
shortest-lived owing to the $W$-exchange contributions, while $\Xi_{bb}^{-}$ and $\Omega_{bb}^{-}$ have similar lifetimes as they both receive contributions from destructive Pauli interference. We find the lifetime ratio $\tau(\Xi_{bb}^{-})/\tau(\Xi_{bb}^0)=1.26$\,.

\begin{table}
\caption{Various contributions to the decay rates (in units of
$10^{-12}$ GeV) of doubly charm-bottom baryons $\B_{bc}$ to order $1/m_b^3$ and $1/m_c^3$ in the HQE with the hadronic scale $\mu_{\rm had}=0.900$ GeV.
}
\label{tab:lifetime3_Bbc}
\begin{center}
\begin{tabular}{l c r c r r c  c  } \hline \hline
 & $\Gamma^{\rm dec}$ & $\Gamma^{\rm ann}$ & $\Gamma^{\rm int}_+$ &
$\Gamma^{\rm int}_-$ &  $\Gamma^{\rm semi}$ & $\Gamma^{\rm tot}$ &
$\tau(10^{-13}s)$~  \\
\hline
 ~$\Xi_{bc}^{+}$~~ & 1.505 &  ~0.034 & $0.060$ & $-0.737$ & ~0.351 &
~1.212~ & ~ 5.43~   \\
 ~$\Xi_{bc}^0$~~ & 1.505 & ~3.807 & ~$0.060$~ & $-0.008$ & ~0.351 &
~5.714~ & ~ 1.15~    \\
 ~$\Omega_{bc}^0$~~ & 1.505 &  & 1.795 & $-0.010$ & ~$0.757$ &
~4.047~ &  ~ 1.63~    \\
\hline \hline
\end{tabular}
\end{center}
\end{table}

For the numerical calculations in the charm-bottom $\B_{bc}$ system, we also need the LO
Wilson coefficients $c_1(\mu)=1.211$ and $c_2(\mu)=-0.430$ at the charm scale $\mu=1.56$ GeV, and the Wilson coefficients $c_1(\mu)=1.172$ and $c_2(\mu)=-0.366$ at $\mu=2m_r=2.32$ GeV.
We use $m_{\Xi_{bc}^*}-m_{\Xi_{bc}}=47$ MeV, $m_{\Omega_{bc}^*}-m_{\Omega_{bc}}=42$ MeV from \cite{Ebert:2005} and the wave function $|\psi^{bc}(0)|^2=0.065\,{\rm GeV}^{3}$ \cite{Baranov}.
The total inclusive rates for the charm-bottom baryons $\B_{bc}$ read
\be
\Gamma(\Xi_{bc}^+) &=& \Gamma^{\rm dec}(\Xi_{bc}^+)+\Gamma^{\rm SL}(\Xi_{bc}^+)+\Gamma^{\B_{bc}}_{{\rm ann},bc}+\Gamma^{\B_{bc}}_{{\rm ann},bu}+\Gamma^{\B_{bc}}_{{\rm int}_+,bc} +\Gamma^{\B_{bc}}_{{\rm int}_-,cu} +\Gamma^{{\rm SL},bc}_{int}, \non \\
\Gamma(\Xi_{bc}^0) &=& \Gamma^{\rm dec}(\Xi_{bc}^0)+\Gamma^{\rm SL}(\Xi_{bc}^0)+\Gamma^{\B_{bc}}_{{\rm ann},bc}+\Gamma^{\B_{bc}}_{{\rm ann},cd}+\Gamma^{\B_{bc}}_{{\rm int}_+,bc} +\Gamma^{\B_{bc}}_{{\rm int}_-,bd} +\Gamma^{{\rm SL},bc}_{int},  \\
\Gamma(\Omega_{bc}^0) &=& \Gamma^{\rm dec}(\Omega_{bc}^0)+\Gamma^{\rm SL}(\Omega_{bc}^0)+\Gamma^{\B_{bc}}_{{\rm ann},bc}+\Gamma^{\B_{bc}}_{{\rm int}_+,bc} +\Gamma^{\B_{bc}}_{{\rm int}_+,cs} +\Gamma^{\B_{bc}}_{{\rm int}_-,bs} +\Gamma^{{\rm SL},bc}_{int}+\Gamma^{{\rm SL},cs}_{int}. \non
\en

The results of calculations to order $1/m_b^3$ amd $1/m_c^3$ are exhibited in Table \ref{tab:lifetime3_Bbc}.
The lifetime hierarchy now reads
\be \label{eq:Bbclifetimepattern}
{\cal O}(1/m_Q^3) \Rightarrow
\tau(\Xi_{bc}^{+})>\tau(\Omega_{bc}^0)>\tau(\Xi_{bc}^0).
\en
Since the CKM matrix element $V_{cs}$ is much larger than $V_{cb}$ in magnitude so that $m_c^2|V_{cs}|^2\gg m_b^2|V_{cb}|^2$, it is obvious that the spectator effects due to $W$-exchange,  constructive and destructive Pauli interferences are dominated by the charm quark, namely $\Gamma^{\B_{bc}}_{{\rm ann},cd}$, $\Gamma^{\B_{bc}}_{{\rm int+},cs}$ and $\Gamma^{\B_{bc}}_{{\rm int-},cu}$, respectively.  Therefore, the large $W$-exchange contribution to $\Xi_{bc}^0$ through the subprocess $cd\to us\to cd$ and the sizable destructive Pauli interference contribution to $\Xi_{bc}^+$ (see Fig. \ref{fig:spectatorbc}) implies a substantial lifetime difference between $\Xi_{bc}^+$ and $\Xi_{bc}^0$. Numerically, we see from Table \ref{tab:lifetime3_Bbc} that $\Gamma^{\B_{bc}}_{{\rm ann},cd}>\Gamma^{\B_{bc}}_{{\rm int+},cs}>|\Gamma^{\B_{bc}}_{{\rm int-},cu}|$. This explains
the lifetime hierarchy  Eq. (\ref{eq:Bbclifetimepattern}). Our lifetime pattern for $\B_{bc}$ baryons to order $1/m_Q^3$ is different from that predicted in \cite{Kiselev:2002,Berezhnoy} (see Table \ref{tab:lifetimes_bc}). Note that $\Omega_{bc}$ has a larger semileptonic rate due to an additional contribution from constructive Pauli interference.

\begin{table}
\caption{Various contributions to the decay rates (in units of
$10^{-12}$ GeV) of doubly charm-bottom baryons $\B_{bc}$ to order $1/m_Q^4$ in the HQE with the hadronic scale $\mu_{\rm had}=0.900$ GeV.
}
\label{tab:lifetime4_Bbc}
\begin{center}
\begin{tabular}{l c r r r r r  c  } \hline \hline
 & $\Gamma^{\rm dec}$ & $\Gamma^{\rm ann}$ & $\Gamma^{\rm int}_+$ &
$\Gamma^{\rm int}_-$ & ~ $\Gamma^{\rm semi}$ & $\Gamma^{\rm tot}$ &
~$\tau(10^{-13}s)$~  \\
\hline
 ~$\Xi_{bc}^{+}$~~ & 1.505 &  ~$-0.044$ & $0.014$ & $0.028$ & ~0.346 &
~1.848 & ~ 3.56~   \\
 ~$\Xi_{bc}^0$~~ & 1.505 & ~5.911 & $0.014$ & $-0.005$ & ~0.346 &
~7.770 & ~ 0.85~    \\
 ~$\Omega_{bc}^0$~~ & 1.505 & $-0.086$ & $-0.759$ & $-0.005$ & ~$0.165$ &
~0.819 &  ~ 8.03~    \\
\hline \hline
\end{tabular}
\end{center}
\end{table}

As shown in \cite{Cheng:2018},
the heavy quark expansion in $1/m_c$ does not work well for describing the lifetime pattern of singly charmed baryons. Since the charm quark is not heavy enough, it is sensible to consider the subleading $1/m_c$ and $1/m_b$ corrections to spectator effects as depicted in Eqs. (\ref{eq:T7baryon}) and (\ref{eq:Spectorbc}). The numerical results are shown in Table \ref{tab:lifetime4_Bbc}. By comparing this table with Table \ref{tab:lifetime3_Bbc}, we see that the lifetimes of $\Xi_{bc}^{+}$ and $\Xi_{bc}^0$ become shorter, whereas $\tau(\Omega_{bc}^0)$ becomes the longest one. The lifetime hierarchy to order $1/m_Q^4$ in the HQE is modified to
$\tau(\Omega_{bc}^{0})>\tau(\Xi_{bc}^+)>\tau(\Xi_{bc}^0)$.
This is similar to the case of singly charmed baryons where the calculated $\Omega_c^0$ lifetime becomes entirely unexpected: the shortest-lived $\Omega_c^0$ to ${\cal O}(1/m_c^4)$ turns out to be the longest-lived one to ${\cal O}(1/m_c^4)$.
This is because $\Gamma^{\rm int}_+$ and $\Gamma^{\rm semi}$ for $\Omega_{bc}^0$ are subject to a large cancellation between dimension-6 and -7 operators.  We see from Table \ref{tab:lifetime4_Bbc} that $\Gamma^{int}_+(\Omega_{bc}^0)$ even becomes negative, while $\Gamma^{int}_-(\Xi_{bc}^+)$ becomes positive. This is because the dimension-7 contribution $\Gamma^{\rm int}_{+,7}(\Omega_{bc}^0)$ is destructive and its size are so large that it overcomes the dimension-6 one and flips the sign. This implies that the subleading corrections are too large to justify the validity of the HQE.

\begin{table}[t]
\caption{Various contributions to the decay rates (in units of
$10^{-12}$ GeV) of the $\Omega_{bc}^0$ after including subleading $1/m_c$ corrections to spectator effects. However, the dimension-7 contributions $\Gamma^{cs}_{{\rm int+,7}}(\Omega_{bc}^0)$, $\Gamma^{{\rm SL},cs}_{\rm int,7}(\Omega_{bc}^0)$, $\Gamma^{cu}_{{\rm int-,7}}(\Xi^+_{bc})$ and $\Gamma^{cd}_{{\rm ann,7}}(\Xi^0_{bc})$ are multiplied by a factor of $(1-\alpha)$ with $\alpha$ varying from 0 to 1. }
\label{tab:lifetime4_charmbary_1}
\begin{center}
\begin{tabular}{c c c c r r l  c |c c} \hline \hline
$\alpha$ & $\Gamma^{\rm dec}$ & $\Gamma^{\rm ann}$ &
$\Gamma^{\rm int}_+$ & $\Gamma^{\rm int}_-$ & ~ $\Gamma^{\rm semi}$ & ~~~~$\Gamma^{\rm tot}$ &
~$\tau(\Omega_{bc}^0)\times 10^{13}$~  & ~~$\Gamma^{\rm int}_-(\Xi_{bc}^+)$ & ~$\Gamma^{cd}_{{\rm ann,7}}(\Xi^0_{bc})$\\
\hline
 ~0~~ & ~~1.505 & ~$-0.086$~ & $-0.759$ & $-0.005$ &  $0.165$ &
~~~0.820~ & ~ 8.03 & ~~0.019 & 5.911  \\
 ~0.31~~ & ~~1.505 & ~$-0.086$~ & ~$0.019$ & $-0.005$ & 0.347 &
~~~1.780~ & ~ 3.70 & $-0.209$  & 5.232   \\
 ~0.45~~ & ~~1.505 & ~$-0.086$~ & ~$0.370$ & $-0.005$ & 0.429 &
~~~2.213~ & ~ 2.97 & $-0.316$  & 4.925 \\
 ~1~~ & ~~1.505 & ~$-0.086$~ & ~$1.750$ & $-0.005$ & 0.752 &
~~~3.916~ & ~ 1.68 & $-0.737$  & 3.721 \\
\hline \hline
\end{tabular}
\end{center}
\end{table}

In order to allow a description of the $1/m_c^4$ corrections to $\Gamma(\B_{bc}^0)$ within the realm of perturbation theory, we follow \cite{Cheng:2018} to introduce a  parameter $\alpha$ so that $\Gamma^{cs}_{{\rm int+,7}}(\Omega_{bc}^0)$, $\Gamma^{{\rm SL},cs}_{\rm int,7}(\Omega_{bc}^0)$, $\Gamma^{cu}_{{\rm int-,7}}(\Xi^+_{bc})$ and $\Gamma^{cd}_{{\rm ann,7}}(\Xi^0_{bc})$ are multiplied by a factor of $(1-\alpha)$; that is, $\alpha$ describes the degree of suppression. In Table \ref{tab:lifetime4_charmbary_1} we show the variation of the $\Omega_{bc}^0$ lifetime with $\alpha$. At $\alpha=0.31$, $\Gamma^{\rm int}_{+}(\Omega_{bc}^0)$ starts to become positive and $\tau(\Omega_{bc}^0)=3.64\times 10^{-13}s$. Since we do not know what the value of $\alpha$ is, we can only conjecture that
it lies in $0.31<\alpha<1$ and
the $\Omega_{bc}^0$ lifetime lies in the range
\be
 1.68\times 10^{-13}s<\tau(\Omega_{bc}^0)< 3.70\times 10^{-13}s.
\en
Likewise,
\be
&& 4.09\times 10^{-13}s<\tau(\Xi_{bc}^+)< 6.07\times 10^{-13}s, \non \\
&& 0.93\times 10^{-13}s<\tau(\Xi_{bc}^0)<1.18\times 10^{-13}s,
\en
with the lifetime pattern $\tau(\Xi_{bc}^+)>\tau(\Omega_{bc}^0)>\tau(\Xi_{bc}^0)$.
The predicted lifetimes of $\B_{bb}$ and $\B_{bc}$ baryons
in this work are summarized in Table \ref{tab:lifetimes_bbbc}.

\begin{table}[t]
\caption{Predicted lifetimes (in units of $10^{-13}s$) of $\B_{bb}$ and $\B_{bc}$ baryons
 in this work.
} \label{tab:lifetimes_bbbc}
\begin{center}
\begin{tabular}{c c c | c c c } \hline \hline
 ~~$\Xi_{bb}^{0}$~~ & ~~$\Xi_{bb}^{-}$~~ & ~~$\Omega_{bb}^{-}$~~ &
 ~~$\Xi_{bc}^{+}$~~ & ~~$\Xi_{bc}^{0}$~~ & ~~$\Omega_{bc}^{0}$~~\\
\hline
6.87 & 8.65 & 8.68 & ~~$4.09\sim 6.07$~ & ~$0.93\sim 1.18$~ & $1.68\sim 3.70$ \\
\hline
\hline
\end{tabular}
\end{center}
\end{table}

\subsection{Comparison with other works}
Comparing Table \ref{tab:lifetimes_bbbc} with Table \ref{tab:lifetimes_bc}, we see that
our lifetime pattern   $\tau(\Omega_{bb}^-)\sim \tau(\Xi_{bb}^{-})>\tau(\Xi_{bb}^0)$ for $\B_{bb}$ baryons is different from the one $\tau(\Xi_{bb}^{-})\approx \tau(\Xi_{bb}^0)$  in \cite{Karliner:2014} and the one $\tau(\Omega_{bb}^-)\approx \tau(\Xi_{bb}^{-})\sim \tau(\Xi_{bb}^0)$  in \cite{Kiselev:2002,Berezhnoy,Likhoded}.
The $\Xi_{bb}^{0}$ baryon is
shortest-lived owing to the $W$-exchange contributions, while $\Xi_{bb}^{-}$ and $\Omega_{bb}^{-}$ have similar lifetimes as they both receive contributions from destructive Pauli interference. We find the lifetime ratio $\tau(\Xi_{bb}^{-})/\tau(\Xi_{bb}^0)=1.26$\,, while it is predicted to be of order unity in \cite{Kiselev:2002,Karliner:2014,Berezhnoy,Likhoded}  (see Table \ref{tab:lifetimes_bc})
due to the smallness of the $W$-exchange in $\Xi_{bb}^0$ and destructive Pauli interference in $\Xi_{bb}^-$. For example, the ratio  of $\Gamma^{\rm ann}(\Xi_{bb}^0)/\Gamma^{\rm dec}$ is calculated to be only one percent (see Table III of \cite{Berezhnoy}), whereas it is of order 0.3 in our case (see Table \ref{tab:lifetime_Bbb}).

For $\B_{bc}$ baryons,  our lifetime hierarchy  $\tau(\Xi_{bc}^+)>\tau(\Omega_{bc}^0)>\tau(\Xi_{bc}^0)$ differs from that of \cite{Kiselev:2002,Berezhnoy,Likhoded} in which
one has $\tau(\Xi_{bc}^+)>\tau(\Xi_{bc}^0)>\tau(\Omega_{bc}^0)$. Since the $bc$ diquark is treated to be a scalar one with $S_{bc}=0$ by the authors of \cite{Kiselev:2002,Berezhnoy,Likhoded}, their $\B_{bc}$ quark matrix elements of four-quark operators and chromomagnetic interactions differ from ours. However, irrespective of the way of treating the $\B_{bc}$ quark matrix elements, an inspection  of Fig. \ref{fig:spectatorbc} leads to  the pattern $\Gamma^{\rm ann}(\Xi_{bc}^0)>\Gamma^{\rm ann}(\Xi_{bc}^+)>\Gamma^{\rm ann}(\Omega_{bc}^0)$. All three $\B_{bc}$ baryons receive a common $W$-exchange contribution through the subprocess $bc\to cs\to bc$, but $\Xi_{bc}^+$ gets an additional $W$-exchange between the $b$ and $u$ quarks, while $\Xi_{bc}^0$ receives a large $W$-exchange contribution through the subprocess $cd\to su\to cd$. Hence, it is not clear to us why $\Omega_{bc}$ has the largest $W$-exchange in \cite{Kiselev:2002,Berezhnoy,Likhoded} (see e.g. Table II of \cite{Berezhnoy}).

As to the Pauli interference, the destructive contribution to $\Xi_{bc}^+$ should be large than the constructive one in magnitude due to the large CKM matrix element $V_{cs}$. As a consequence, the net Pauli interference in $\Xi_{bc}^+$ should be negative, while it was calculated to be positive in \cite{Kiselev:2002,Berezhnoy,Likhoded}.

\section{Conclusions}
In this work we have analyzed the lifetimes of the doubly heavy baryons $\B_{bb}$ and $\B_{bc}$ within the framework of the heavy quark expansion. It is well known that the lifetime differences stem from spectator effects such as $W$-exchange and Pauli interference.  We rely on the quark model to evaluate the hadronic matrix elements of dimension-6 and -7 four-quark operators responsible for spectator effects.

\vskip 0.2cm
The main results of our analysis are as follows.
\begin{itemize}
\item A special attention is paid to the doubly heavy baryon matrix elements of dimension-3 and -5 operators which are different from the ones of singly heavy baryons.
The doubly doubly baryon matrix element of the $\sigma\cdot G$ operator receives three distinct contributions: the interaction of the heavy quark with the chromomagnetic field produced from the light quark and from the other heavy quark, and the so-called Darwin term in which the heavy quark interacts with the chromoelectric field. For $\B_{bc}$ baryons, the first contribution is proportional to the hyperfine mass splitting of $\B_{bc}$. However, it vanishes if the $bc$ diquark is wrongly assigned to be of the scalar type as often assumed in the previous studies.

\item For doubly bottom baryons, the lifetime pattern is  $\tau(\Omega_{bb}^-)\sim \tau(\Xi_{bb}^{-})>\tau(\Xi_{bb}^0)$.
The $\Xi_{bb}^{0}$ baryon is
shortest-lived owing to the $W$-exchange contributions, while $\Xi_{bb}^{-}$ and $\Omega_{bb}^{-}$ have similar lifetimes as they both receive contributions from destructive Pauli interference. We find the lifetime ratio $\tau(\Xi_{bb}^{-})/\tau(\Xi_{bb}^0)=1.26$\,.

\item The study of $\B_{bc}$ lifetimes is more complicated than the $\B_{bb}$ case for several reasons. First, besides the spectator effects due to each heavy quark $b$ or $c$, there also exist $W$-exchange and Pauli interference in which both $b$ and $c$ quarks get involved. Second, care must be taken when considering the heavy quark expansion for the charm quark.

\item The large $W$-exchange contribution to $\Xi_{bc}^0$ through the subprocess $cd\to us\to cd$ and the sizable destructive Pauli interference contribution to $\Xi_{bc}^+$ imply a substantial lifetime difference between $\Xi_{bc}^+$ and $\Xi_{bc}^0$.

\item  In the presence of subleading $1/m_c$ and $1/m_b$ corrections to the spectator effects, we find that $\tau(\Omega_{bc}^0)$ becomes longest-lived. This is because $\Gamma^{\rm int}_+$ and $\Gamma^{\rm semi}$ for $\Omega_{bc}^0$ are subject to large cancellation between dimension-6 and -7 operators.  This implies that the subleading corrections are too large to justify the validity of the HQE. Demanding that $\Gamma^{cs}_{{\rm int+}}(\Omega_{bc}^0)$, $\Gamma^{{\rm SL},cs}_{\rm int}(\Omega_{bc}^0)$ be positive and $\Gamma^{cu}_{{\rm int-}}(\Xi^+_{bc})$ be negative, we conjecture that $1.68\times 10^{-13}s<\tau(\Omega_{bc}^0)< 3.70\times 10^{-13}s$,  $4.09\times 10^{-13}s<\tau(\Xi_{bc}^+)< 6.07\times 10^{-13}s$ and $0.93\times 10^{-13}s<\tau(\Xi_{bc}^0)< 1.18\times 10^{-13}s$.

\item The lifetime hierarchy in the $\B_{bc}$ system is expected to be $\tau(\Xi_{bc}^{+})>\tau(\Omega_{bc}^0)>\tau(\Xi_{bc}^0)$. We have compared our work with others.

\end{itemize}

\section{Acknowledgments}

This research  was supported in part by the Ministry of Science and Technology of R.O.C. under Grant No. 107-2119-M-001-034.  F. Xu was supported by NSFC under Grant No. 11605076 as well as the Fundamental Research Funds for the Central Universities in China under the Grant No. 21616309.

%
%
\newcommand{\bi}{\bibitem}
%

\end{document}